\documentclass[aps,twocolumn,superscriptaddress,showpacs,showkeys,preprintnumbers,amsmath,amssymb,amsfonts,nofootinbib]{revtex4}

\pdfoutput=1

\usepackage{graphicx}
\usepackage{color}
\usepackage{slashed}
\usepackage{tabularx}
\usepackage{hyperref}

\newcommand{\Tr}{\mathrm{Tr}}

\newcommand{\I}{\mathrm{i}}
\newcommand{\Nf}{N_{\text{f}}}
\newcommand{\Nc}{N_{\text{c}}}

\newcolumntype{L}[1]{>{\raggedright\arraybackslash}p{#1}} %linksbündig mit Breitenangabe
\newcolumntype{C}[1]{>{\centering\arraybackslash}p{#1}} %zentriert mit Breitenangabe
\newcolumntype{R}[1]{>{\raggedleft\arraybackslash}p{#1}} %rechtsbündig mit Breitenangabe

\graphicspath{{.}{figures/}{mma/}}

\begin{document}

\title{Flow equations for spectral functions at finite external momenta}

\newcommand{\TUD}{Theoriezentrum, Institut f\"ur Kernphysik, Technische Universit\"at Darmstadt, 64289 Darmstadt, Germany}
\newcommand{\JLU}{Institut f\"ur Theoretische Physik, Justus-Liebig-Universit\"at Giessen, 35392 Giessen, Germany}
\newcommand{\GSI}{GSI Helmholtzzentrum f\"ur Schwerionenforschung GmbH, 64291 Darmstadt, Germany}

\author{Ralf-Arno Tripolt}\affiliation{\TUD}
\author{Lorenz von Smekal}\affiliation{\TUD}\affiliation{\JLU}
\author{Jochen Wambach}\affiliation{\TUD}\affiliation{\GSI}

\begin{abstract}
In this work we study the spatial-momentum dependence of mesonic spectral functions obtained from the quark-meson model using a recently proposed method to calculate real-time observables at finite temperature and density from the Functional Renormalization Group. This non-perturbative method is thermodynamically consistent, symmetry-preserving and based on an analytic continuation from imaginary to real time on the level of the flow equations for 2-point functions. Results on the spatial-momentum dependence of the pion and sigma spectral function are presented at different temperatures and densities, in particular near the critical endpoint in the phase diagram of the quark-meson model.
\end{abstract}

\pacs{12.38.Aw, 12.38.Lg, 11.10.Wx, 11.30.Rd}
\keywords{spectral function, analytic continuation, QCD phase diagram, chiral phase transition}

%12.38.Aw: general properties of QCD (dynamics, confinement, etc.)
%12.38.Lg: other nonperturbative calculations
%11.10.Wx: finite-temperature field theory
%11.30.Rd: chiral symmetries

\maketitle

\section{Introduction}

The in-medium properties of hadrons are of particular interest to strong interaction physics, not least due to currently running and future heavy-ion collision experiments. The interpretation of dilepton spectra measured at various invariant masses and transverse momenta in these experiments, for example, requires  a detailed theoretical understanding of the spatial-momentum dependence of in-medium vector spectral functions \cite{Braun-Munzinger2009,Friman:2011zz}.  

The calculation of real-time quantities like spectral functions or transport coefficients represents, however, a great challenge within Euclidean approaches to thermal quantum field theory since one has to perform an analytic continuation from imaginary to real time \cite{Baym1961,Landsman1987}. This technical difficulty arises for example in Lattice QCD, where numeric data at discrete imaginary Matsubara frequencies has to be used to reconstruct real-time correlations using techniques like the maximum entropy method (MEM) \cite{Jarrell:1996,Asakawa:2000tr,Ding:2012sp}. When based on such numerical schemes, the analytic continuation from Euclidean to Minkowski space-time is, however, an ill-posed inverse problem. 

Therefore, any approach that allows for a well-defined analytic continuation procedure is highly desirable. Such alternative methods were proposed for example in \cite{Kamikado2013, Kamikado2014, Tripolt2014} and \cite{Floerchinger2012} for the Functional Renormalization Group (FRG) and involve analytic continuations on the level of the flow equations. The FRG represents a powerful continuum framework for non-perturbative calculations particularly in quantum field theory and statistical physics, see \cite{Berges:2000ew,Pawlowski:2005xe,Schaefer:2006sr,Braun:2011pp,Gies2012} for reviews. In addition to providing a complementary approach to, e.g., Monte Carlo simulations on the lattice or Dyson-Schwinger Equations (DSEs), the FRG allows for a proper inclusion of quantum fluctuations which is particularly well suited to describe critical phenomena, for example. 

In the following we will adopt the FRG approach developed in our previous work \cite{Tripolt2014}, whose perhaps most appealing aspect is its simplicity while it nevertheless has several important properties built in. In particular it is symmetry preserving, i.e.\ chiral symmetry and its breaking pattern are realized exactly. Moreover, it is thermodynamically consistent, i.e.\ the space-like limit of zero external momentum in the 2-point functions agrees with the curvature masses as extracted from the thermodynamic grand potential \cite{Kamikado2014,Tripolt2014}. Furthermore, the method satisfies the physical Baym-Mermin boundary conditions \cite{Baym1961} which are implemented essentially as in a simple one-loop calculation \cite{Lebellac:1996,Das:1997gg}. In addition, it can be systematically improved towards full QCD \cite{Haas:2013qwp,Herbst:2013ufa} and to calculate quark and gluonic spectral functions as an alternative to analytically continued DSEs \cite{Strauss:2012dg} or to using MEM on Euclidean FRG \cite{Haas:2013hpa} or DSE results \cite{Nickel:2006mm,Mueller:2010ah,Qin:2013ufa}. 

In this work we extend the method presented in \cite{Tripolt2014} to include arbitrary external spatial momenta for 2-point functions and use it to calculate in-medium mesonic spectral functions. In particular, results are shown for the pion and sigma spectral function obtained by employing the quark-meson model \cite{Jungnickel:1995fp,Schaefer:2004en}, which is used as an effective description of QCD at low energies. The calculation of transport coefficients like the shear viscosity \cite{Aarts:2002cc,Haas:2013hpa}, using the momentum dependent in-medium spectral functions as input, represents an interesting application of our approach and will be the subject of upcoming work.

This paper is organized as follows. In Sec.~\ref{sec:setup} we briefly introduce our theoretical setup, i.e.\ the FRG approach to the quark-meson model, the analytic continuation procedure and the numerical implementation. In Sec.~\ref{sec:results} we discuss the possible decay channels and processes affecting the spectral functions and present results on the spatial-momentum dependence of the pion and sigma spectral function at different temperatures and quark chemical potentials, in particular near the critical endpoint in the phase diagram of the quark-meson model. A summary and an outlook are given in Sec.~\ref{sec:summary}. The appendices contain explicit expressions for threshold and loop functions, and assessments of the extent of Lorentz symmetry breaking induced by our truncation and of the parameter dependence of the results.

\section{Theoretical Setup}\label{sec:setup}
In this section we briefly outline the employed FRG approach to the quark-meson model, derive flow equations for mesonic 2-point functions at finite external spatial momentum and discuss our analytic continuation procedure. For a more detailed discussion, in particular on the numerical implementation, we refer to \cite{Tripolt2014}.

\subsection{Functional renormalization group and quark-meson model}\label{sec:FRG_QM}
\begin{figure}[b]
\includegraphics[width=0.65\columnwidth]{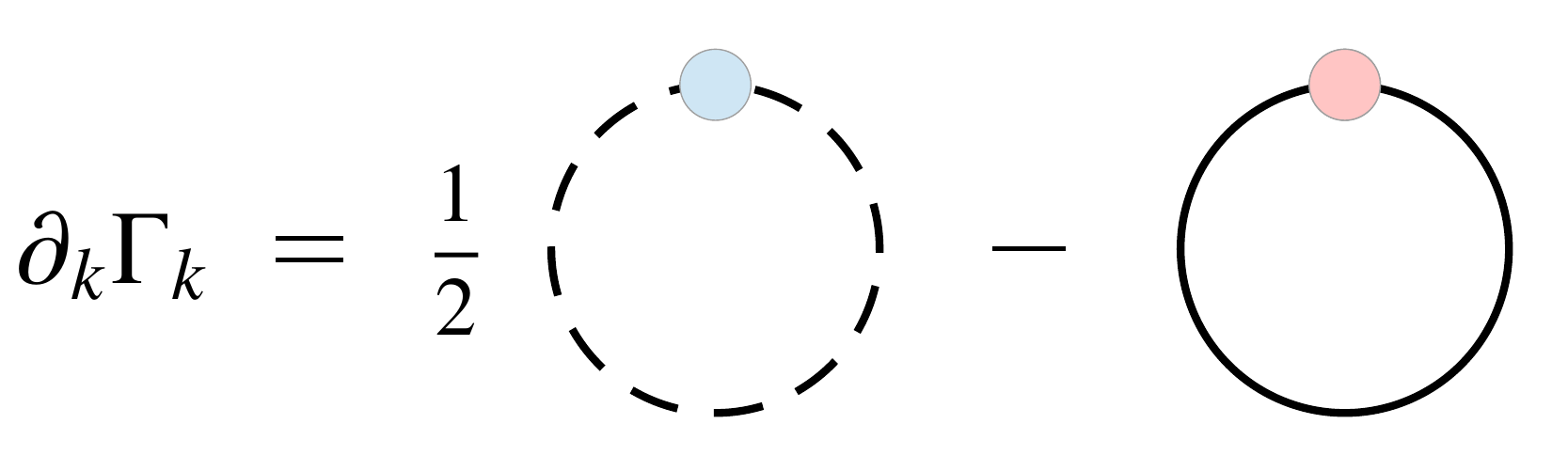}
\caption{(color online) Diagrammatic representation of the flow equation for the effective average action. Dashed (solid) lines represent bosonic (fermionic) propagators and blue (red) circles represent bosonic (fermionic) regulator insertions~$\partial_k R_k^{B,F}$.}
\label{fig:flow_Gamma} 
\end{figure}

The basic idea of the FRG is to take quantum fluctuations into account successively by introducing an infrared (IR) regulator function $R_k$, where the associated RG scale $k$ has to be lowered from some ultraviolet (UV) cutoff scale down to zero. The scale dependence of the effective average action, $\Gamma_k$, which becomes the usual quantum effective action $\Gamma$ in the limit $k\rightarrow 0$, is given by the following exact flow equation \cite{Wetterich:1992yh, Morris1994}, also known as the Wetterich equation,
\begin{equation}
\label{eq:wetterich}
\partial_k \Gamma_k=\frac{1}{2}\text{STr}\left\{ \partial_k R_k\left(\Gamma^{(2)}_k+R_k\right)^{-1}\right\} ,
\end{equation}
where $\Gamma^{(2)}_k$ denotes the second functional derivative of the effective average action, and the supertrace includes a summation over internal and space-time indices as well as an integration over the loop momentum. A graphical representation of this equation is given by Fig.~\ref{fig:flow_Gamma} for bosonic and fermionic fields.

\begin{figure*}[t]
\includegraphics[width=1.7\columnwidth]{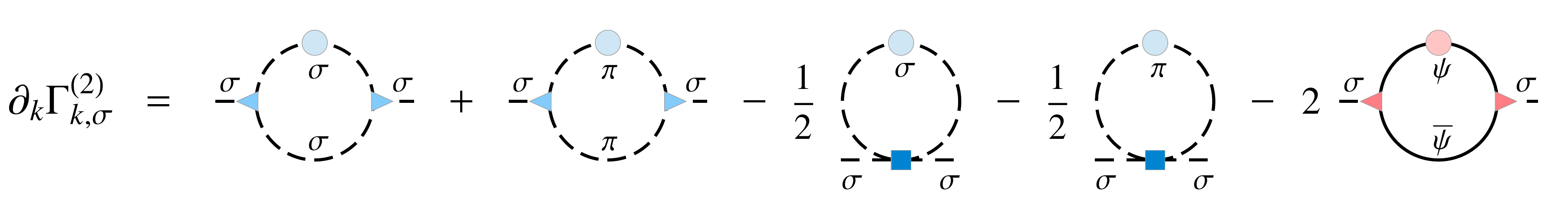}\\
\includegraphics[width=1.7\columnwidth]{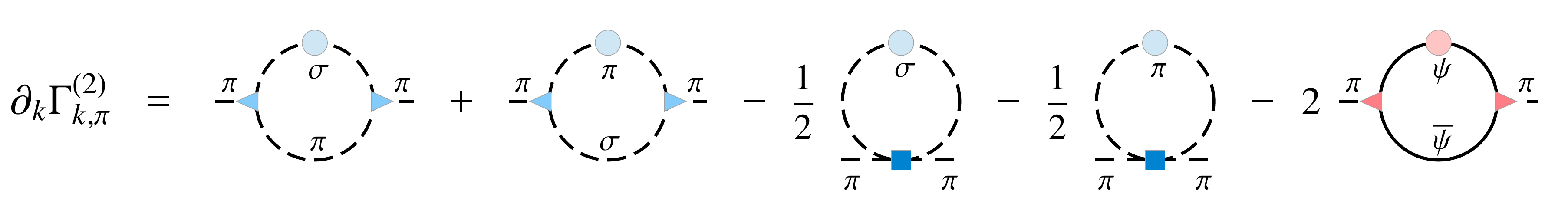}
\caption{(color online) Diagrammatic representation of the flow equation for the mesonic 2-point functions. Mesonic and quark-meson three-point vertices are represented by blue and red triangles respectively, mesonic four-point vertices by blue squares. The bosonic and fermionic regulator insertions are depicted by blue and red circles respectively.}
\label{fig:flow_Gamma2}
\end{figure*}

The Wetterich equation is here applied to the quark-meson model, which serves as an effective model for QCD with two light quark flavors at low energies \cite{Jungnickel:1995fp, Schaefer:2004en}. In particular, it realizes the chiral symmetry breaking pattern and allows to go beyond the mean-field approximation by including fluctuations due to collective mesonic excitations. We use the following truncation for the RG-scale dependent effective average action, also known as local potential approximation (LPA),
\begin{eqnarray} 
\Gamma_{k}[\bar \psi,\psi,\phi]&=& 
\int d^{4}x \:\Big\{
\bar{\psi} \left({\partial}\!\!\!\slash +
h(\sigma+i\vec{\tau}\cdot\vec{\pi}\gamma_{5}) -\mu \gamma_0 \right)\psi\nonumber \\
&& \qquad\hspace{4mm}
+\frac{1}{2} (\partial_{\mu}\phi)^{2}+U_{k}(\phi^2) - c \sigma 
\Big\}\,,
\label{eq:QM}
\end{eqnarray}
where $\psi$ represents $N_\text{f}=2$ quark flavors, the mesonic fields are combined as $\phi\equiv(\sigma,\vec{\pi})$
and a term $c\sigma$, which is equivalent to a quark mass term, is added in order to achieve explicit chiral symmetry breaking.
The effective potential $U_{k}$ is chosen to be of the form 
\begin{equation}
\label{eq:pot_UV} 
U_\Lambda(\phi^{2}) =
\frac{1}{2}m_\Lambda^{2}\phi^{2} +
\frac{1}{4}\lambda_\Lambda(\phi^{2})^{2}
\end{equation}
at the UV scale $\Lambda$ but is allowed to develop non-vanishing higher order couplings when fluctuations are taken into account at lower scales. The flow equation for the effective potential is obtained by combining Eqs.~(\ref{eq:wetterich}) and (\ref{eq:QM}) and reads
\begin{equation}
\label{eq:flow_pot} 
\partial_k U_k =\frac{1}{2} I_{k,\sigma}^{(1)} +\frac{3}{2} I_{k,\pi}^{(1)} -\Nc \Nf I_{k,\psi}^{(1)},
\end{equation}
where threshold functions $I_{k,\alpha}^{(n)}$ are defined as
\begin{equation}
\label{eq:I_def} 
I_{k,\alpha}^{(n)} \equiv \Tr_q \left[ \partial_k R^{B,F}_k(q)\,G_{k,\alpha}^{n}(q)\right],
\end{equation}
with $\alpha\in\{\sigma,\pi, \psi\}$ and, for later, powers $n=1,2$ of the Euclidean propagators $G_{k,\alpha}(q)$, i.e.
\begin{equation}
\label{eq:propagator} 
G^n_{k,\alpha}(q) \equiv \left( \Gamma_{k,\alpha}^{(2)}(q)+R^{B,F}_k(q)\right)^{-n},
\end{equation}
where the regulator function is chosen appropriately for bosonic (B) and fermionic (F) fields respectively. The trace in (\ref{eq:I_def}) includes both a momentum integration and a Dirac trace for fermionic expressions. Explicit expressions for the regulator functions are given by Eqs.~(\ref{eq:3dregulators}) and (\ref{eq:3dregulators2}) in Sec.~\ref{sec:analytic_continuation}, while for explicit expressions of the threshold functions we refer to App.~\ref{app:thresholds}.

In order to calculate spectral functions, we first have to derive flow equations for the 2-point functions, which are obtained by taking two functional field-derivatives of Eq.~(\ref{eq:wetterich}) and can be expressed as
\begin{align}
\label{eq:gamma2sigma}
\partial_k \Gamma^{(2)}_{k,\sigma}(p)&=\partial_k \Gamma^{(2),B}_{k,\sigma}(p)+
\partial_k \Gamma^{(2),F}_{k,\sigma}(p),\\
\partial_k \Gamma^{(2)}_{k,\pi}(p)&=\partial_k \Gamma^{(2),B}_{k,\pi}(p)+
\partial_k \Gamma^{(2),F}_{k,\pi}(p),
\label{eq:gamma2pion}
\end{align}
where $p$ is the external 4-momentum and the bosonic and fermionic contributions are given by \cite{Tripolt2014}
\begin{align}
\label{eq:Gamma2sigmaB}
\partial_k \Gamma^{(2),B}_{k,\sigma}(p)=&\:
J_{k,\sigma\sigma}(p)(\Gamma_{k,\sigma\sigma\sigma}^{(0,3)})^2
+3\, J_{k,\pi\pi}(p) (\Gamma_{k,\sigma\pi\pi}^{(0,3)})^2 \nonumber \\
&-\frac{1}{2}I_{k,\sigma}^{(2)}\Gamma_{k,\sigma\sigma\sigma\sigma}^{(0,4)}
-\frac{3}{2} I_{k,\pi}^{(2)}\Gamma_{k,\sigma\sigma\pi\pi}^{(0,4)}\,,\\
\label{eq:Gamma2pionB}
\partial_k \Gamma^{(2),B}_{k,\pi}(p)=&\:
J_{k,\sigma\pi}(p) (\Gamma_{k,\sigma\pi\pi}^{(0,3)})^2+
J_{k,\pi\sigma}(p)(\Gamma_{k,\sigma\pi\pi}^{(0,3)})^2\nonumber \\
&-\frac{1}{2}I_{k,\sigma}^{(2)}\Gamma_{k,\sigma\sigma\pi\pi}^{(0,4)}
-\frac{5}{2}I_{k,\pi}^{(2)}\Gamma_{k,\pi\pi\tilde{\pi}\tilde{\pi}}^{(0,4)}\,,\\
\label{eq:Gamma2sigmaF}
\partial_k \Gamma^{(2),F}_{k,\sigma}(p)=&-2\Nc \Nf \,J^{(\sigma)}_{k,\bar{\psi}\psi}(p)\,,\\
\label{eq:Gamma2pionF}
\partial_k \Gamma^{(2),F}_{k,\pi}(p)=&-2\Nc \Nf\, J^{(\pi)}_{k,\bar{\psi}\psi}(p)\,,
\end{align}
with $\pi,\tilde{\pi}\in \{\pi_1,\pi_2,\pi_3\}$ and $\pi\neq\tilde{\pi}$. A diagrammatic representation of the flow equations for the pion and sigma two-point functions is given by Fig.~\ref{fig:flow_Gamma2}. In the following we choose the external momentum to be routed through the ``upper" leg of the loop diagrams in Fig.~\ref{fig:flow_Gamma2}, i.e.\ the one with the regulator insertion $\partial_k R_k$. This choice is convenient but of course arbitrary and we have verified explicitly that our results do not depend on the momentum routing. The loop functions $J_{k,\alpha\beta}(p)$ and $J^{(\alpha)}_{k,\bar{\psi}\psi}(p)$ are then given by 
\begin{align}
\label{eq:J_def}
J_{k,\alpha\beta}(p)=\:&\Tr_q
\left[ 
\partial_k R^B_{k}(q+p)G_{k,\alpha}(q+p)^2G_{k,\beta}(q)
\right], \\
\label{eq:J_F_def}
J^{(\alpha)}_{k,\bar{\psi}\psi}(p)=\:&\Tr_q 
\Big[ 
\partial_k R^F_{k}(q+p)G_{k,\psi}(q+p)
\Gamma_{\bar{\psi}\psi\alpha}^{(2,1)}\nonumber \\
& \left.\qquad G_{k,\psi}(q)\Gamma_{\bar{\psi}\psi\alpha}^{(2,1)}G_{k,\psi}(q+p)
\right],
\end{align}
with $\alpha,\beta\in\{ \sigma,\pi\}$. For explicit expressions of the analytically continued loop functions we refer to App.~\ref{app:thresholds}.

The mesonic three and four-point vertex functions are extracted from the effective potential $U_k(\phi^2)$ and are therefore scale dependent but momentum independent. Explicitly, they are given by
\begin{eqnarray}
\label{eq:meson_vertex_3}
\Gamma^{(0,3)}_{k,\phi_i\phi_j\phi_m}&=& 4 U''_k \left(  \delta_{ij}\phi_m + 
\delta_{im}\phi_j + \delta_{jm}\phi_i \right)\nonumber\\
&&+\,8U^{(3)}_k \phi_i\phi_j\phi_m
\,,\\[3mm]
\label{eq:meson_vertex_4}
\Gamma^{(0,4)}_{k,\phi_i\phi_j\phi_m\phi_n}&=&4 U''_k 
\left(  \delta_{ij}\delta_{mn} + \delta_{in}\delta_{jm} + \delta_{jn}\delta_{im} 
\right) \nonumber\\
&&+\,
8U^{(3)}_k \left( \delta_{ij}\phi_m\phi_n + \delta_{jm}\phi_i\phi_n + 
\delta_{mn}\phi_i\phi_j \right. \nonumber \\
&& \left. \qquad \quad + \,\delta_{jn}\phi_i\phi_m + 
\delta_{in}\phi_j\phi_m + \delta_{im}\phi_j\phi_n  \right) \nonumber\\
&&+\,
16 U^{(4)}_k \phi_i\phi_j\phi_m\phi_n\,,
\end{eqnarray}
where $U_k^{(n)}$ represents the $n$-th derivative with respect to $\phi^2$. Obtaining all mesonic vertices and higher $n$-point functions from the effective potential has the particular advantage that our truncation is thermodynamically consistent, in the sense that the mesonic 2-point functions agree with the corresponding meson masses as extracted from the effective potential in the space-like limit of zero external momentum \cite{Kamikado2014,Tripolt2014}. 

Finally, the quark-meson three-point vertex functions are basically given by the Yukawa coupling $h$, which is taken to be momentum and scale independent in the present truncation, i.e.
\begin{equation}
\label{eq:quark_vertex}
\Gamma^{(2,1)}_{\bar \psi \psi  \phi_i}=h \begin{cases}1 &\text{for}\: \:i=0\\
\I\gamma^5\tau^i &\text{for}\:\: i=1,2,3\,. \end{cases}
\end{equation}

\subsection{Analytic continuation}\label{sec:analytic_continuation}

Our aim in this section is to analytically continue the flow equations for the mesonic 2-point functions, Eqs.~(\ref{eq:gamma2sigma}) and (\ref{eq:gamma2pion}), from imaginary to real frequencies. For that, we use the method described in \cite{Tripolt2014} which is based on a simple one-loop calculation in thermal field theory \cite{Lebellac:1996,Das:1997gg} and can be applied here since the flow equations for the 2-point functions do also have a one-loop structure. Moreover, it is necessary to perform the internal frequency integration, or Matsubara summation at finite temperature, on the right-hand side of the flow equations explicitly. We therefore choose three-dimensional regulator functions which do not affect the frequency component of the internal 4-momentum.\footnote{For a discussion on the possible violation of Lorentz invariance induced by the 3d regulator functions we refer to App.~\ref{app:lorentz}.} Explicitly, we employ the following 3d analogues of the LPA-optimized regulator functions \cite{Litim:2001up}, 
\begin{align}
\label{eq:3dregulators}
R^B_{k}(q)&=(k^2-\vec q^{\,2})\Theta(k^2-\vec q^{\,2})\,,\\
R^F_{k}(q)&=i \slashed{\vec q} (\sqrt{k^2/\vec q^{\,2}}-1)
\Theta(k^2-\vec q^{\,2})\,.
\label{eq:3dregulators2}
\end{align}
The flow equations for the 2-point functions can then be represented in terms of bosonic and fermionic occupation numbers, cf.\ App.~\ref{app:thresholds}, which allows for the 
following 2-step analytic continuation procedure.

We first exploit the periodicity of the bosonic and fermionic occupation numbers with respect to the discrete external (mesonic) Euclidean $p_0$ in thermal equilibrium, which is given by multiples of $2\pi T$, i.e. 
\begin{equation}
\label{eq:continuation1}
n_{B,F}(E+\I p_0)\rightarrow n_{B,F}(E).
\end{equation}
As a second step, the flow equations for the retarded 2-point functions are obtained from their Euclidean counterparts by replacing the discrete Euclidean energy $p_0$ by a continuous real frequency $\omega$ in the following way,
\begin{equation}
\label{eq:continuation2}
\Gamma^{(2),R}(\omega,\vec p)=-\lim_{\epsilon\to 0} \Gamma^{(2),E}(p_0=-\I(\omega+\I\epsilon), \vec p).
\end{equation}
where $\epsilon$ is taken to be a small real parameter in our numerical implementation, namely $\epsilon=1$~MeV. We note that these two steps have to be carried out in the given order so as to recover the physical Baym-Mermin boundary conditions \cite{Baym1961}. For explicit expressions of the analytically continued threshold and loop functions appearing in the flow equations for the retarded 2-point functions we refer to App.~\ref{app:thresholds}.

Finally, the spectral functions are given by the negative of the imaginary part of the propagators, which can be expressed in terms of the retarded 2-point functions as
\begin{equation}
\label{eq:spectralfunction}
\rho(\omega,\vec p)=\frac{1}{\pi}\frac{\text{Im}\,\Gamma^{(2),R}(\omega,\vec p)}{\left(\text{Re}\,
\Gamma^{(2),R}(\omega,\vec p)\right)^2+\left(\text{Im}\,\Gamma^{(2),R}(\omega,\vec p)\right)^2}.
\end{equation}

\subsection{Numerical implementation}\label{sec:numeric_implementation}

In order to solve the flow equations for the retarded 2-point functions we first solve the flow equation for the effective potential, Eq.~(\ref{eq:flow_pot}). This is done by discretizing the radial field component $\sigma$ along a one-dimensional grid in field space while the angular field components are set to their expectation value, $\langle \vec{\pi}\rangle=0$. In this way the original partial differential equation turns into a set of ordinary differential equations which can be solved numerically \cite{Schaefer:2004en}.
In the following we choose the parameters of the quark-meson model such as to reproduce phenomenologically reasonable values for the quark and meson masses in the IR, cf.\ Table~\ref{tab:parameters}. Explicitly, we use UV parameters such that at  an IR scale of $k_\mathrm{IR}=40$~MeV,\footnote{In practice, the integration of the flow equation for the effective potential has to be stopped some sufficiently small but finite RG~scale $k_\mathrm{IR}$, cf.~\cite{Schaefer:2004en}.} one obtains the following values for the vacuum curvature masses of the mesons, the quark mass and the pion decay constant, which is identified with the global minimum of the effective potential: $m_\pi=138$~MeV, $m_\sigma=509$~MeV, $m_\psi=299$~MeV and $f_\pi\equiv \sigma_0=93.5$~MeV.
Moreover, one can estimate the resulting quark condensate from the trace of the quark propagator subtracted at the UV cutoff $\Lambda$ \cite{Smekal1991}.
With our parameters this yields for the vacuum condensate at $T=\mu=0$, for example,  $\langle \bar{\psi}\psi\rangle\approx (257\text{ MeV})^3$, in good agreement with QCD sum rules and consistent with the GOR relation. Its temperature and density dependence is essentially that of $\sigma_0$ in this simple model.
\begin{table}[b]
\centering
\begin{tabular}{C{1.3cm}|C{1.1cm}|C{1.5cm}|C{1.1cm}}
$m_\Lambda/\Lambda$ & $\lambda_\Lambda$ & $c/\Lambda^3$ &  $h$  \\
\hline
0.794 & 2.00 & 0.00175 & 3.2
\end{tabular}
\caption{Parameter set for the quark-meson model at an UV scale of $\Lambda=1000$~MeV.}
\label{tab:parameters} 
\end{table}

The flow equations for the retarded 2-point functions, cf.\ Eqs.~(\ref{eq:gamma2sigma}) and (\ref{eq:gamma2pion}), are also solved by using the grid method and by using the scale-dependent effective potential as input. Since the flow equations for the 2-point functions do not couple to other values of the 2-point functions at different values of the $\sigma$ field, the grid in field space reduces to only one point, i.e.\ $\sigma=\sigma_0$. As initial condition for the retarded 2-point functions at the UV scale, we use
\begin{eqnarray}
\label{eq:UV_sigma} 
\Gamma^{(2),R}_{\Lambda,\sigma}(\omega, \vec{p})&=&\omega^2-\vec{p}^{\,2}-2U_{\Lambda}'-4\phi^2 U_{\Lambda}'', \\
\label{eq:UV_pion} 
\Gamma^{(2),R}_{\Lambda,\pi}(\omega, \vec{p})&=&\omega^2-\vec{p}^{\,2}-2U_{\Lambda}',
\end{eqnarray}
where $\omega$ has to be replaced by $(\omega+\I\epsilon)$. The dependence of our results on the value of $\epsilon$ is studied in App.~\ref{app:epsilon} while for further details on the numerical implementation we refer to \cite{Tripolt2014}.

\section{Results}\label{sec:results}

In this section we first discuss the possible quasi-particle processes that our framework allows for and then present results on the momentum dependence of the sigma and pion spectral functions.

\subsection{Available decay channels and quasi-particle processes}\label{sec:processes}
First, we distinguish between time-like and space-like processes, i.e.\ those contributing for external 4-momenta $p\equiv (\omega,\vec p)$ with $p^2=\omega^2-\vec{p}^{\,2}>0$ and $p^2<0$. In general, time-like processes can be interpreted in terms of particle creation, annihilation and absorption reactions, while space-like processes represent the emission or absorption of space-like excitations.

\begin{figure*}[ht]
\includegraphics[width=0.32\columnwidth]{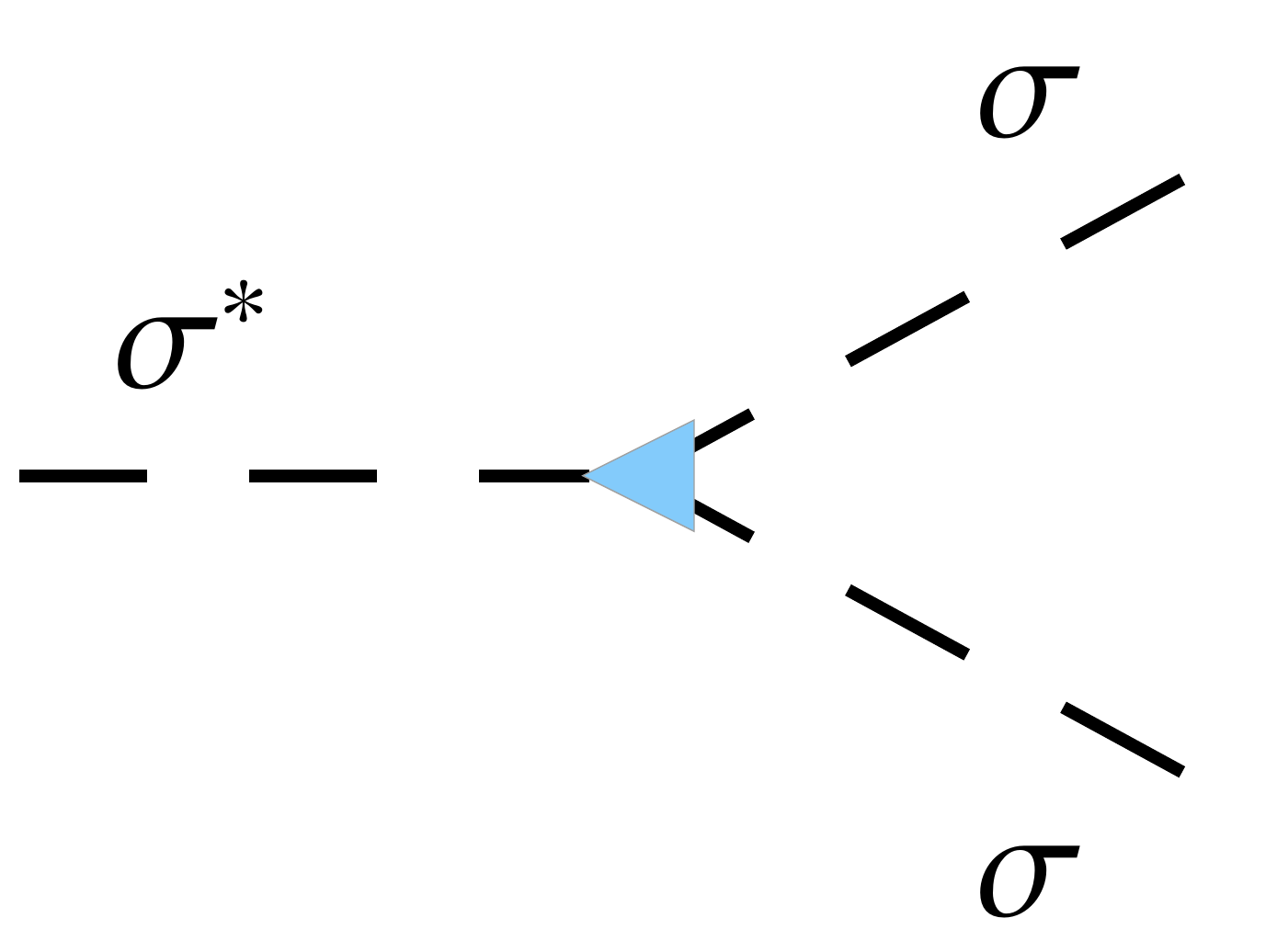}\hspace{15mm}
\includegraphics[width=0.32\columnwidth]{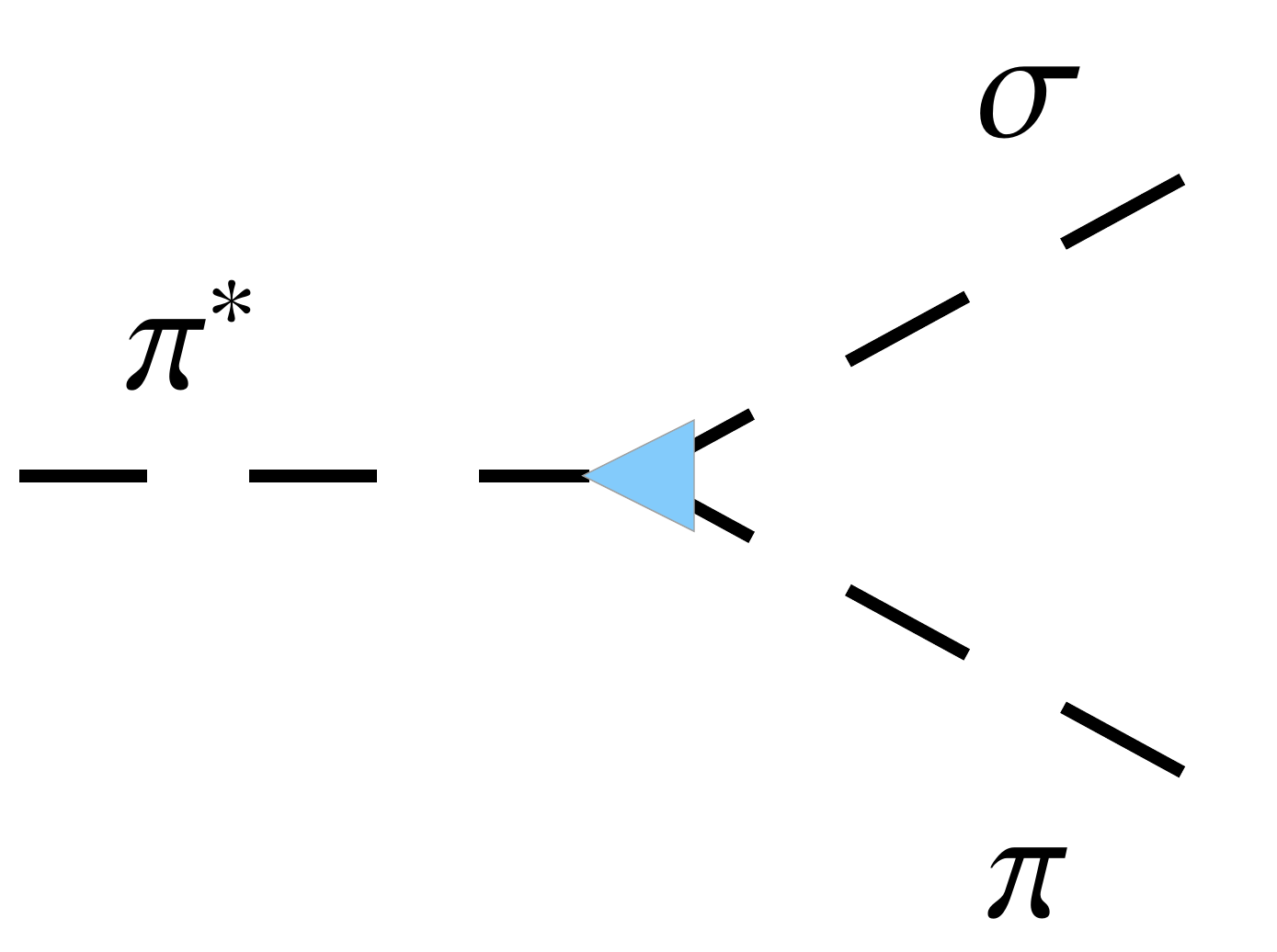}\hspace{15mm}
\includegraphics[width=0.32\columnwidth]{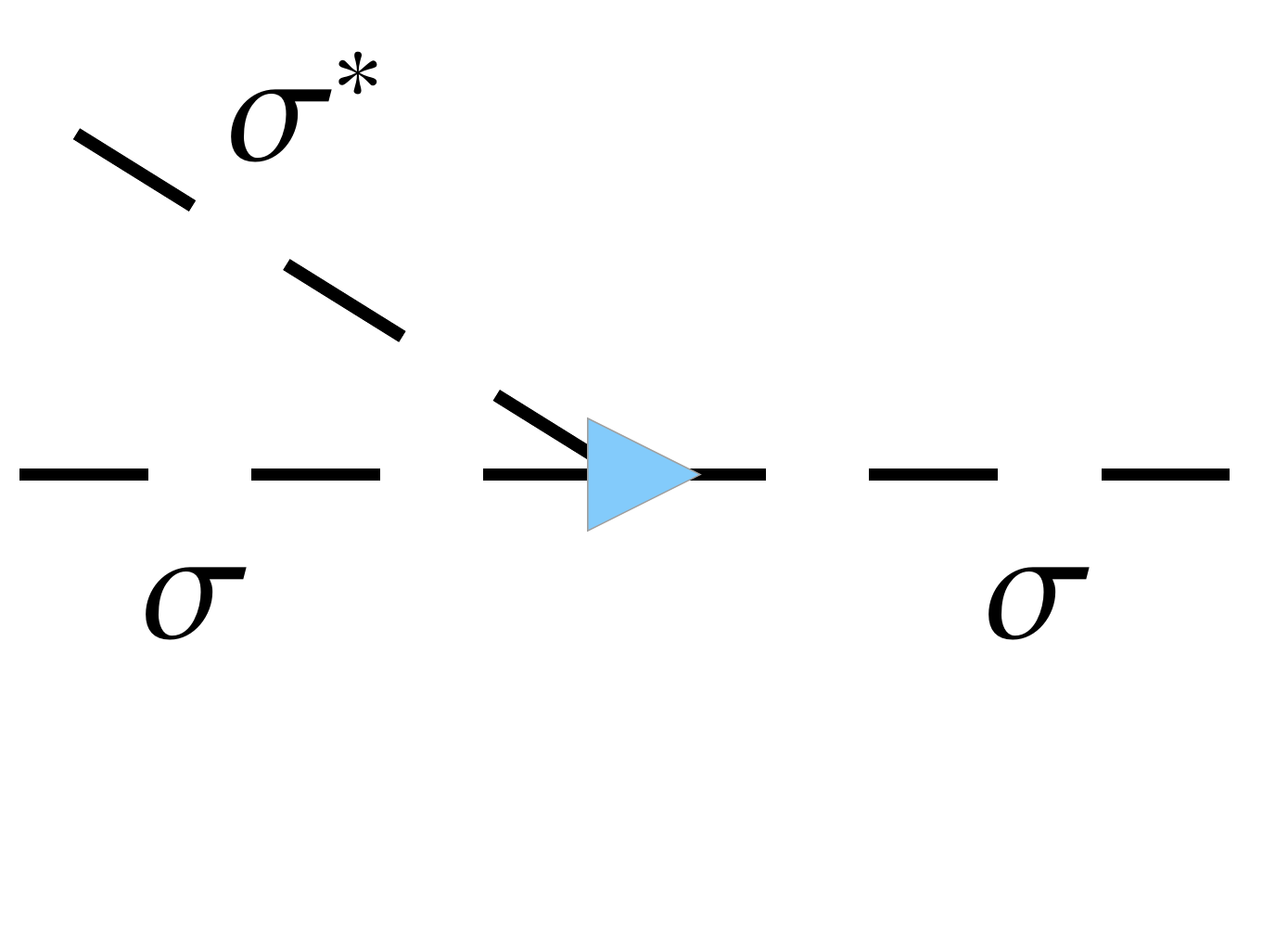}\hspace{15mm}
\includegraphics[width=0.32\columnwidth]{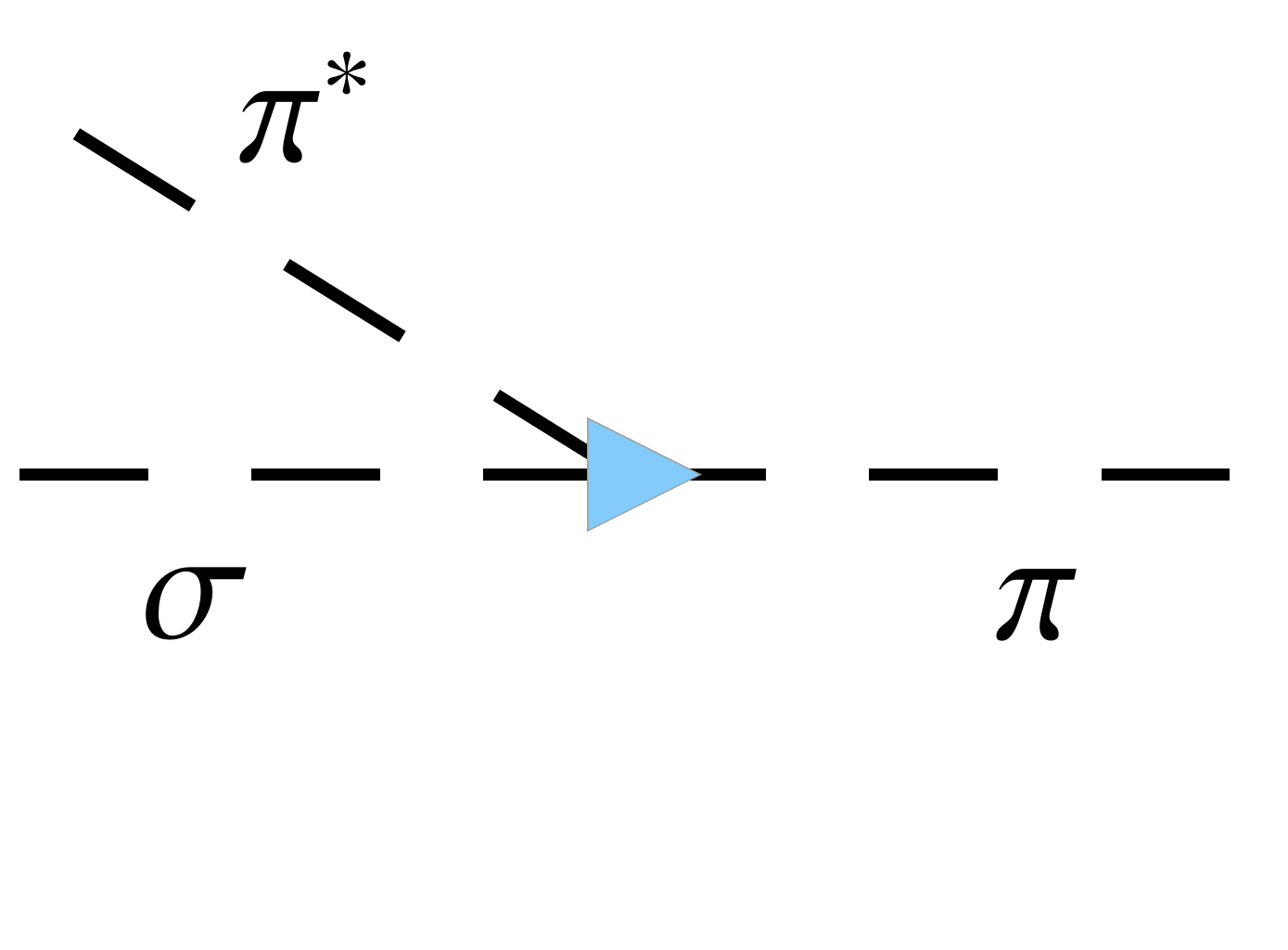}\\[3mm]
\includegraphics[width=0.32\columnwidth]{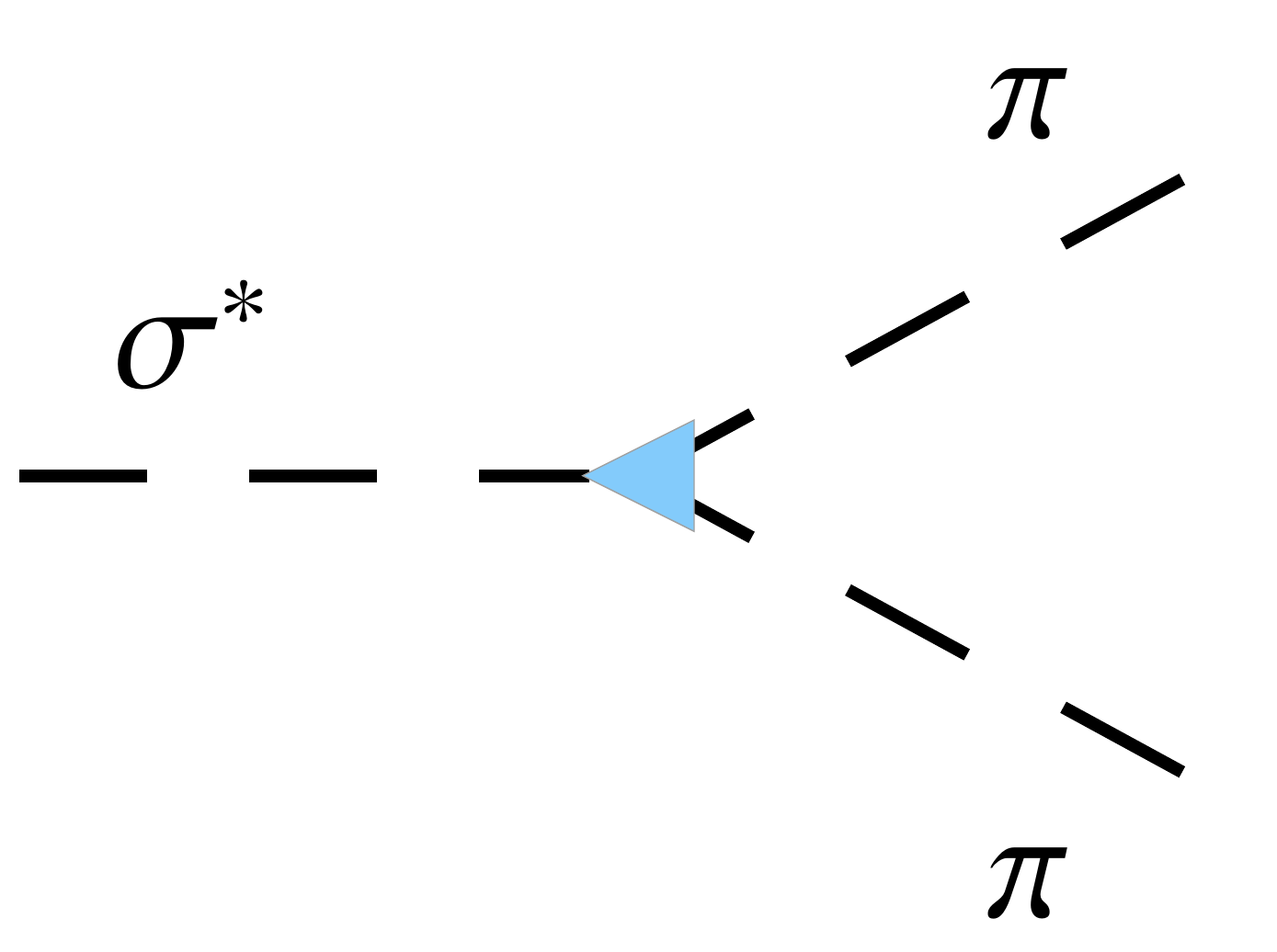}\hspace{15mm}
\includegraphics[width=0.32\columnwidth]{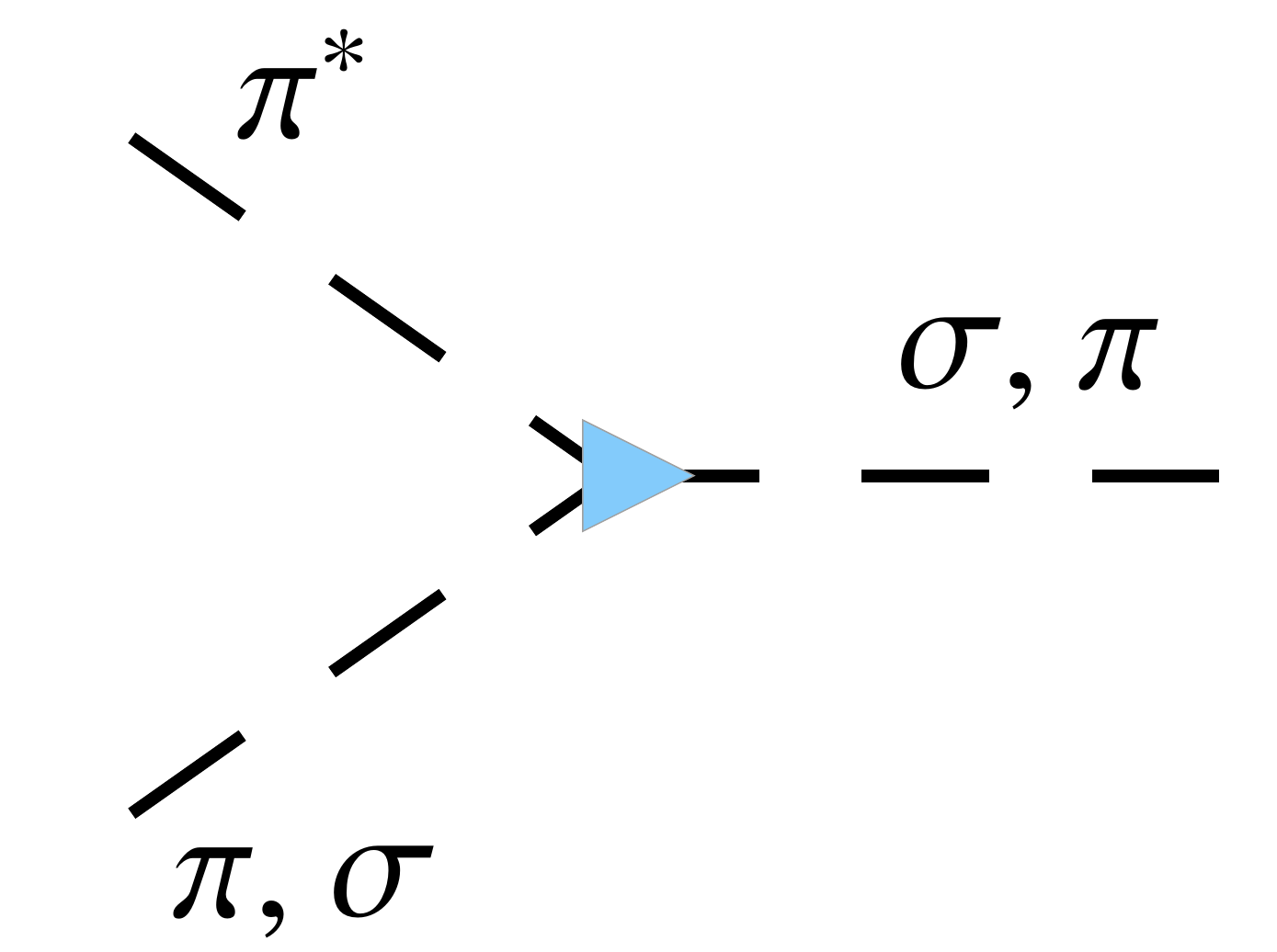}\hspace{15mm}
\includegraphics[width=0.32\columnwidth]{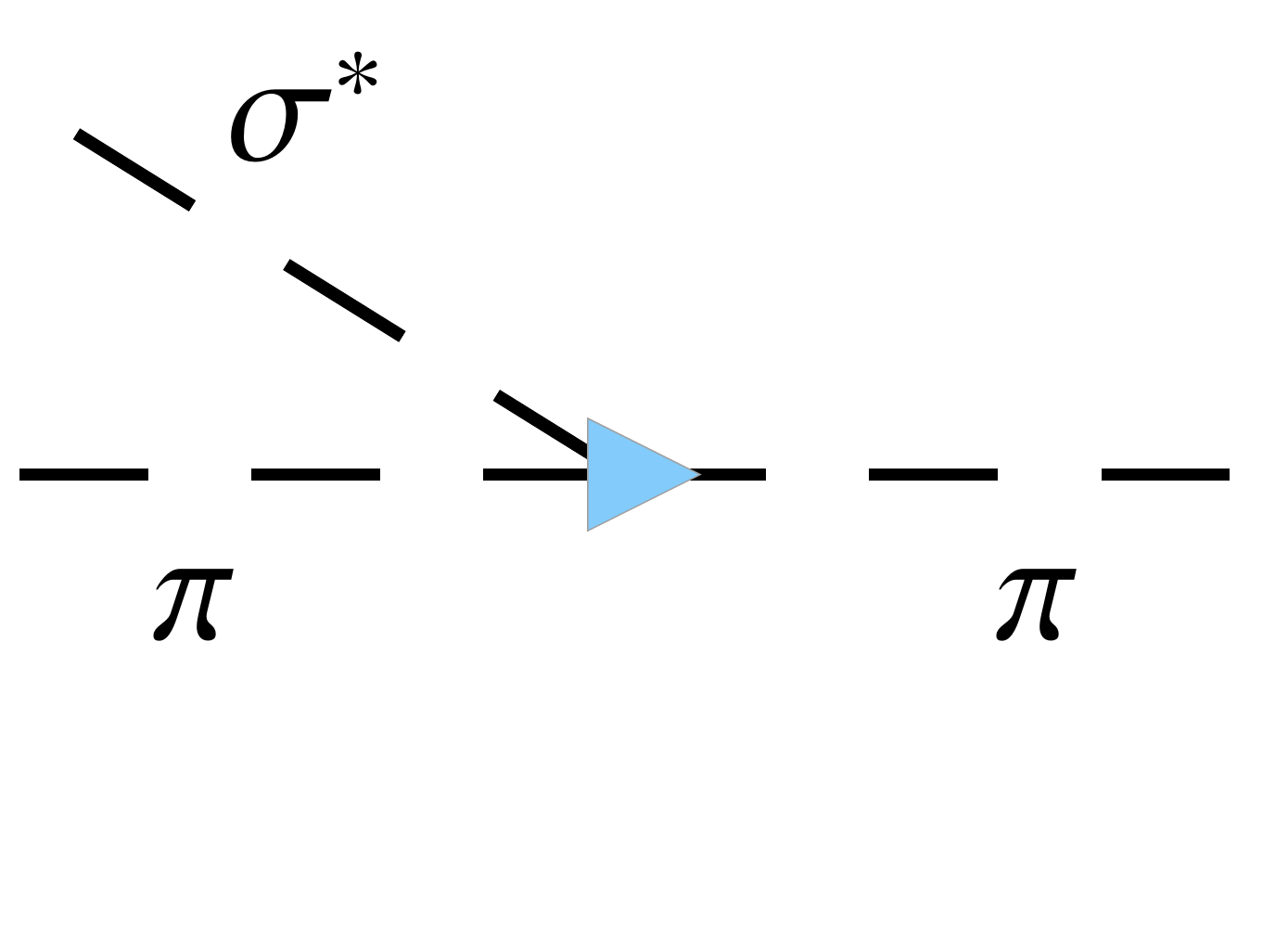}\hspace{15mm}
\includegraphics[width=0.32\columnwidth]{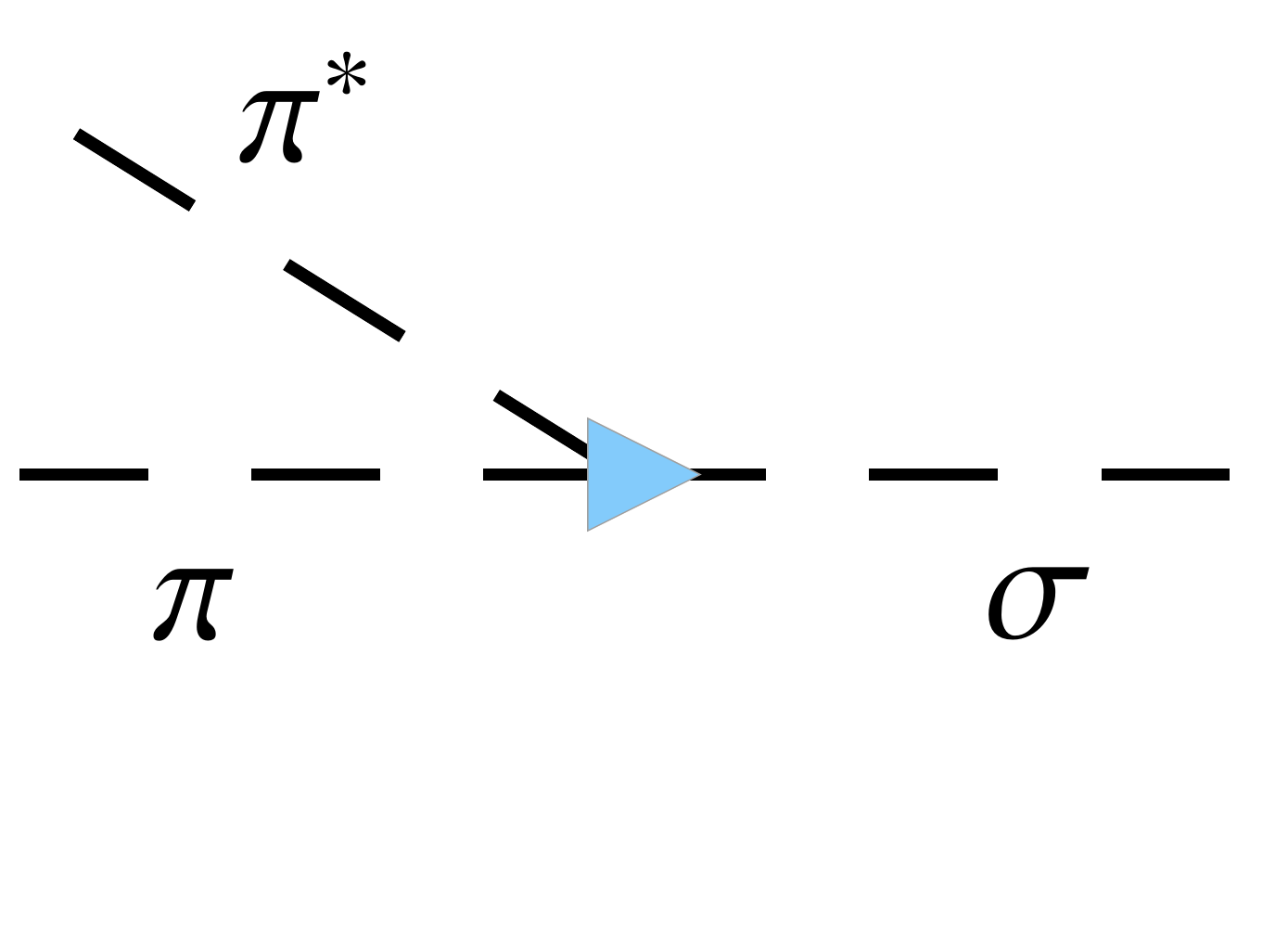}\\[3mm]
\includegraphics[width=0.32\columnwidth]{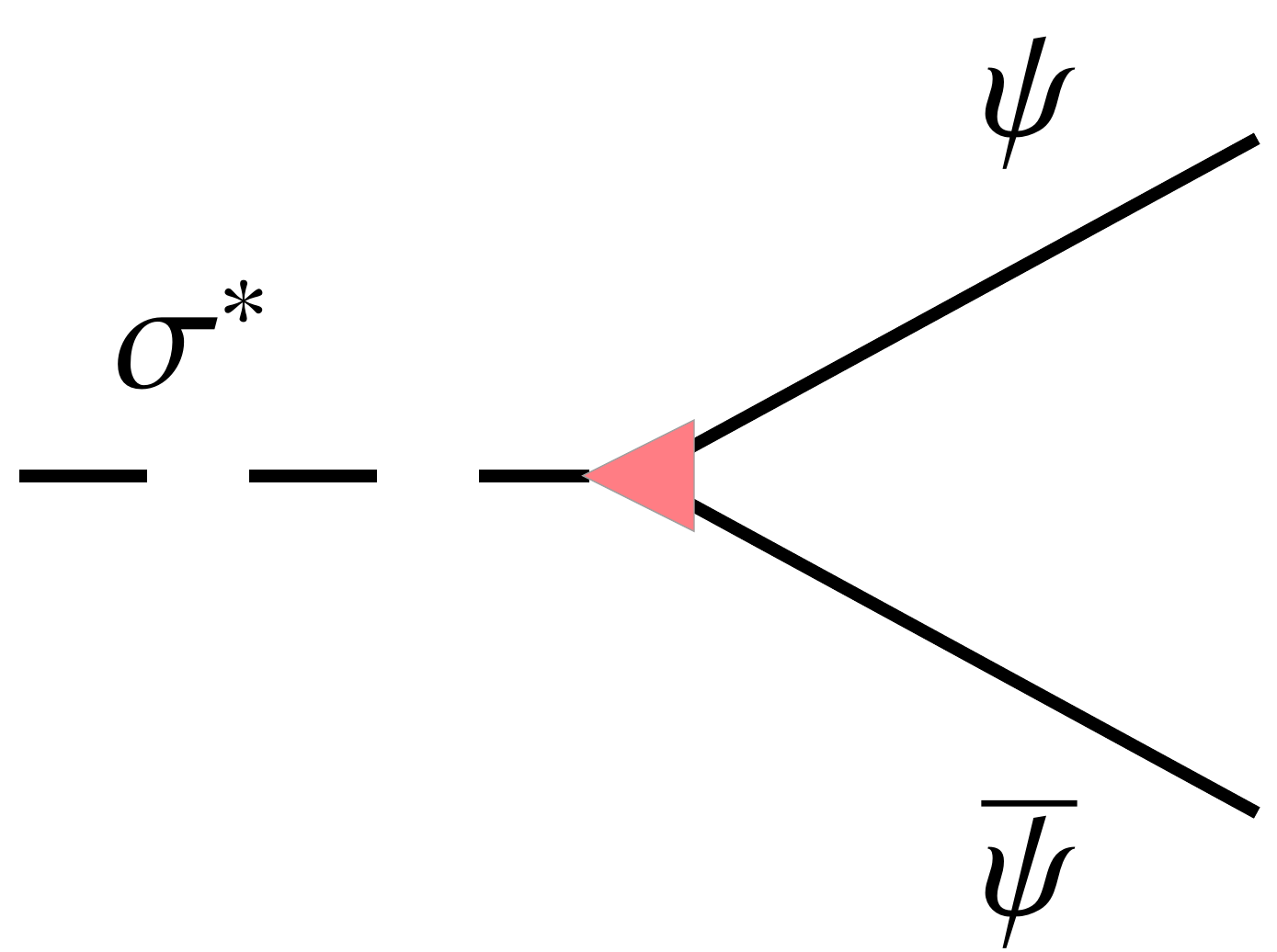}\hspace{15mm}
\includegraphics[width=0.32\columnwidth]{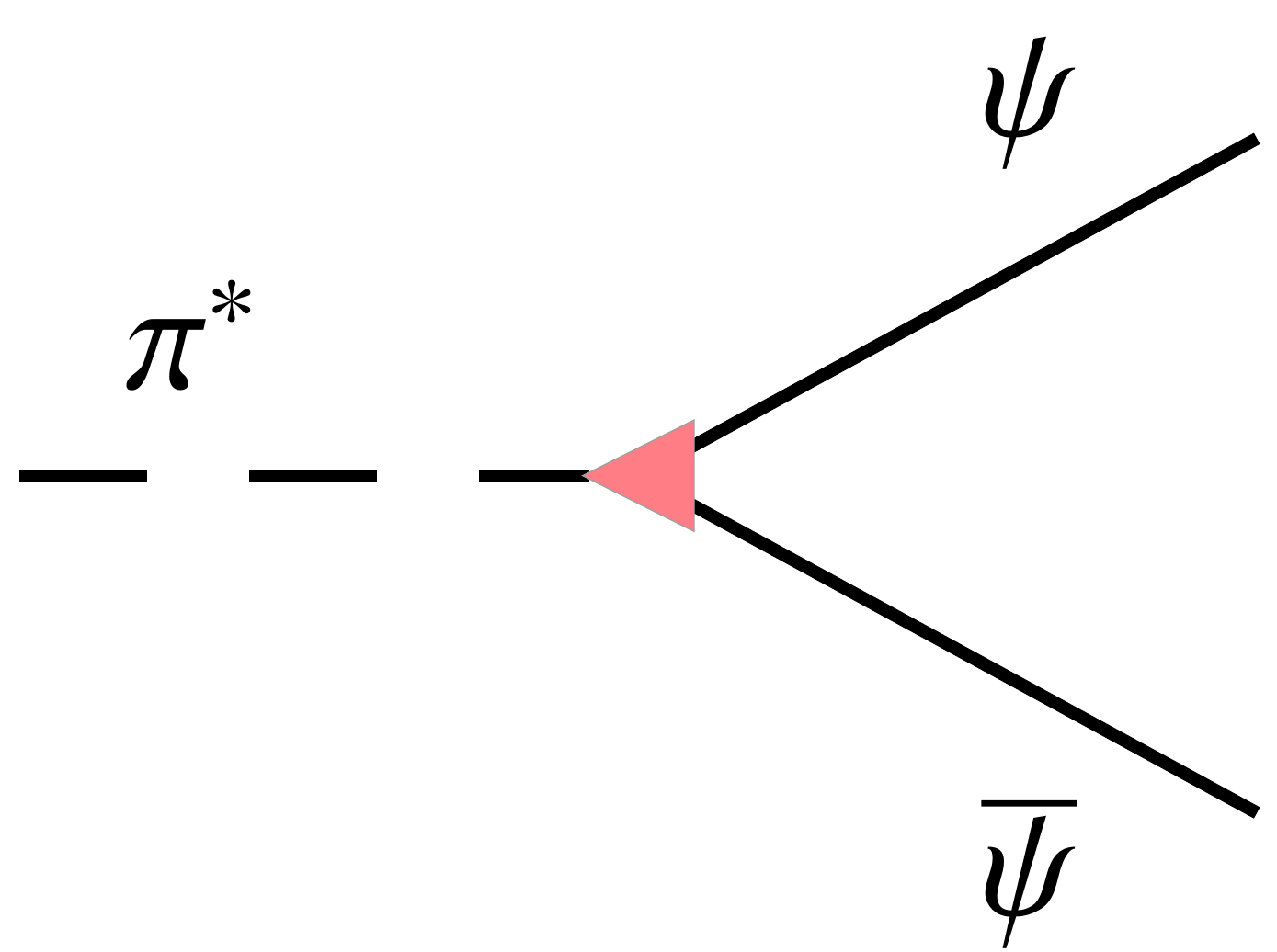}\hspace{15mm}
\includegraphics[width=0.32\columnwidth]{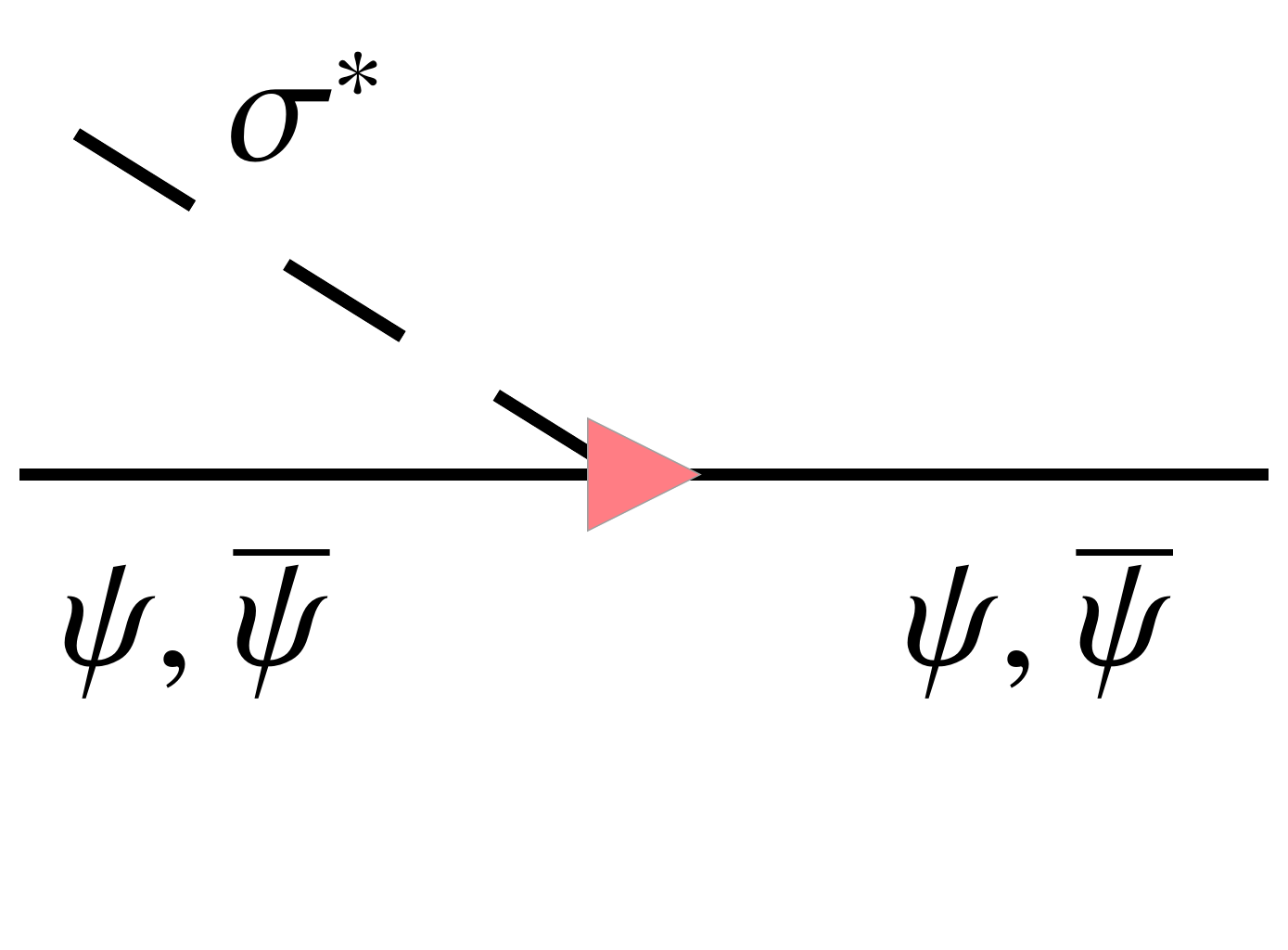}\hspace{15mm}
\includegraphics[width=0.32\columnwidth]{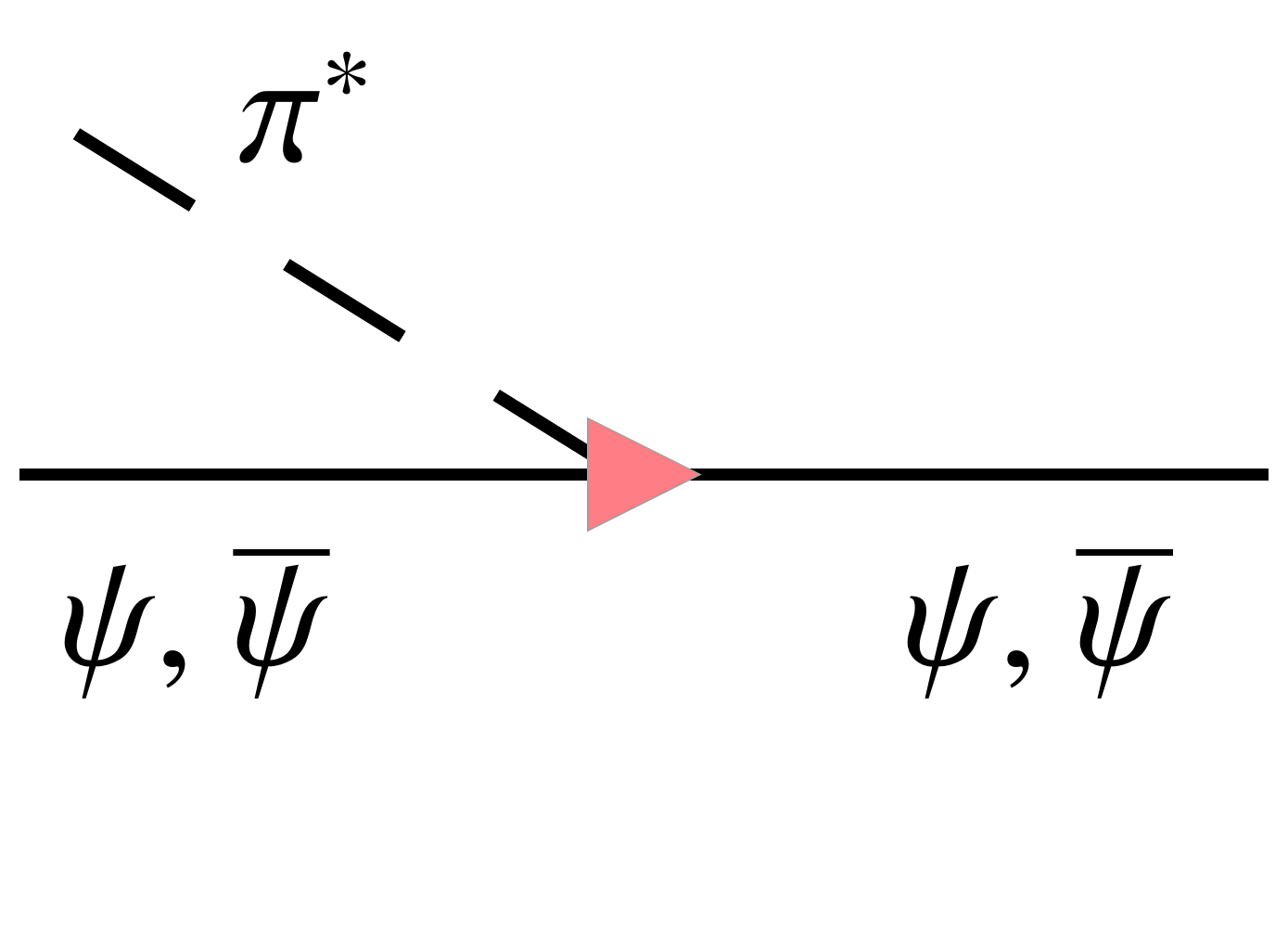}
\caption{(color online) Graphical representation of the possible time-like, $p^2=\omega^2-\vec{p}^{\,2}>0$, and space-like, $p^2<0$, processes. Asterisks denote off-shell particles while others represent on-shell particles from a heat bath. First column: time-like decay channels for an off-shell sigma, $\sigma^*$. Second column: time-like decay channels for an off-shell pion, $\pi^*$. Third column: absorption or emission of a space-like sigma excitation. Fourth column: absorption or emission of a space-like pion excitation.}
\label{fig:processes} 
\end{figure*}

The available time-like processes for an external off-shell sigma, denoted as $\sigma^*$, with energy $\omega$ and spatial momentum $\vec{p}$ are given by
\begin{align}
\label{eq:sigma_sigma_sigma_decay}
&\sigma^*\rightarrow \sigma + \sigma\,, & \omega \geq \sqrt{(2m_\sigma)^2+\vec{p}^{\,2}},\\
\label{eq:sigma_pion_pion_decay}
&\sigma^*\rightarrow \pi + \pi\,,& \omega \geq \sqrt{(2m_\pi)^2+\vec{p}^{\,2}},\\
\label{eq:sigma_quark_quark_decay}
&\sigma^*\rightarrow \bar{\psi} + \psi\,,&  \omega \geq \sqrt{(2m_\psi)^2+\vec{p}^{\,2}},
\end{align}
where the kinematic constraints follow from energy conservation. These processes are already evident from Fig.~\ref{fig:flow_Gamma2} and are represented explicitly in Fig.~\ref{fig:processes}. We note that the decay products are on-shell, since they represent real particles in a heat bath, and that their masses are to be identified with the curvature masses in our truncation. Moreover, the inverse of the above processes is also possible at finite temperature, giving rise to an equilibrium between direct and inverse processes. In fact, the total decay rates for these processes can be readily obtained from the flow equations for the retarded 2-point functions, which can be expressed in terms of statistical weight factors, i.e.\ combinations of occupation numbers, cf.\ App.~\ref{app:thresholds}.

The time-like processes for an off-shell pion $\pi^*$ are given by
\begin{align}
\label{eq:pion_sigma_pion_decay}
&\pi^*\rightarrow \sigma + \pi\,, & \omega \geq \sqrt{(m_\sigma+m_\pi)^2+\vec{p}^{\,2}},\\
\label{eq:pion_pion_sigma_decay_T}
&\pi^*+\pi \rightarrow \sigma\,,& |\vec{p}|\leq \omega \leq (m_\sigma-m_\pi)\sqrt{1+\tfrac{\vec{p}^{\,2}}{\Delta m^2}},\\
\label{eq:pion_sigma_pion_decay_T}
&\pi^*+\sigma \rightarrow \pi\,,& |\vec{p}|\leq \omega \leq (m_\pi-m_\sigma)\sqrt{1+\tfrac{\vec{p}^{\,2}}{\Delta m^2}},\\
\label{eq:pion_quark_quark_decay}
&\pi^*\rightarrow \bar{\psi} + \psi\,,&  \omega \geq \sqrt{(2m_\psi)^2+\vec{p}^{\,2}},
\end{align}
with $\Delta m^2\equiv (m_\sigma-m_\pi)^2$, cf.\ Fig.~\ref{fig:processes}. We note that only either the absorption process given by Eq.~(\ref{eq:pion_pion_sigma_decay_T}) or that by Eq.~(\ref{eq:pion_sigma_pion_decay_T}) is possible. As they require real particles from the heat bath they are furthermore only possible at finite temperature in either case. Because the sigma mass is usually the larger one,  for most combinations of $T$ and $\mu$ only the process $\pi^*+\pi \rightarrow \sigma$ can occur. It is replaced by the process $\pi^*+\sigma \rightarrow \pi$ only in a small region around the critical endpoint where $m_\sigma<m_\pi$, cf.\ Sec.~\ref{sec:results_p}. Moreover, the lower kinematic bound of these absorption processes, $|\vec{p}|\leq \omega$, only arises due to their classification as being  time-like. In this case, however, the same processes can also occur for space-like external momentum configurations as will be discussed below. The distinction between time-like and space-like is therefore somewhat artificial for these absorption processes.

The possible space-like processes for a sigma excitation are given by
\begin{equation}
\label{eq:spacelike_sigma}
\left.
\begin{aligned}
&\sigma^* + \sigma \rightarrow \sigma\qquad\\
&\sigma^* + \pi \rightarrow \pi\qquad\\
&\sigma^* + \psi \rightarrow \psi\qquad
\end{aligned}
\right\}
\qquad 0\leq \omega \leq |\vec{p}|
\end{equation}
where $\psi$ represents both the quark and anti-quark field, cf.\ Fig.~\ref{fig:processes}. They describe the absorption or, when considering the inverse processes, the emission of an off-shell sigma. 

Similarly, the space-like processes for a pion excitation are
\begin{equation}
\label{eq:spacelike_pion}
\left.
\begin{aligned}
&\pi^* + \sigma \rightarrow \pi\qquad\\
&\pi^* + \pi \rightarrow \sigma\qquad\\
&\pi^* + \psi \rightarrow \psi\qquad
\end{aligned}
\right\}
\qquad 0\leq \omega \leq |\vec{p}|
\end{equation}
which are also represented in Fig.~\ref{fig:processes}. We note that all space-like processes require real particles from the heat bath and are therefore only possible at finite temperature or density.

\subsection{Momentum dependence of spectral functions}\label{sec:results_p}
As an instructive example, that allows to study the effects of most of the different time-like and space-like processes discussed in the previous section, we first demonstrate the spatial-momentum dependence of the sigma and pion spectral functions at a temperature of $T=100$~MeV for vanishing quark chemical potential. We then present corresponding results with non-zero external spatial momenta near the critical endpoint (CEP) in the phase diagram of the quark-meson model. For our choice of parameters, cf.\ Tab.~\ref{tab:parameters}, the chiral crossover occurs at temperatures of around $170$~MeV at zero chemical potential, while the CEP is located approximately at $T=10$~MeV and $\mu=293$~MeV, cf.\ \cite{Tripolt2014} for a more detailed discussion of the phase diagram and for results on the temperature and chemical potential dependence of the spectral functions at $|\vec{p}|=0$. 

\begin{figure*}[t]
\includegraphics[width=\columnwidth]{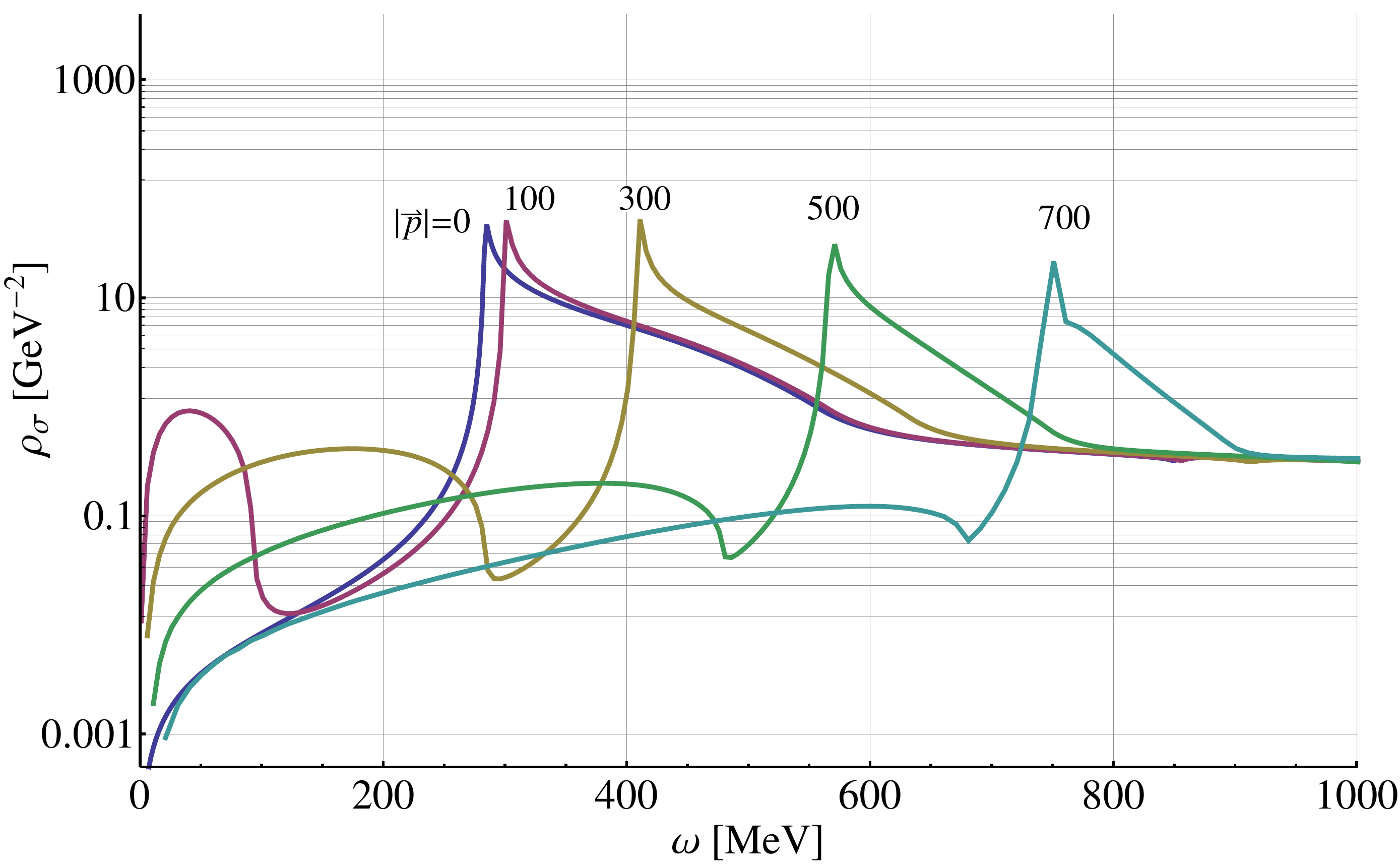}\hspace{3mm}
\includegraphics[width=\columnwidth]{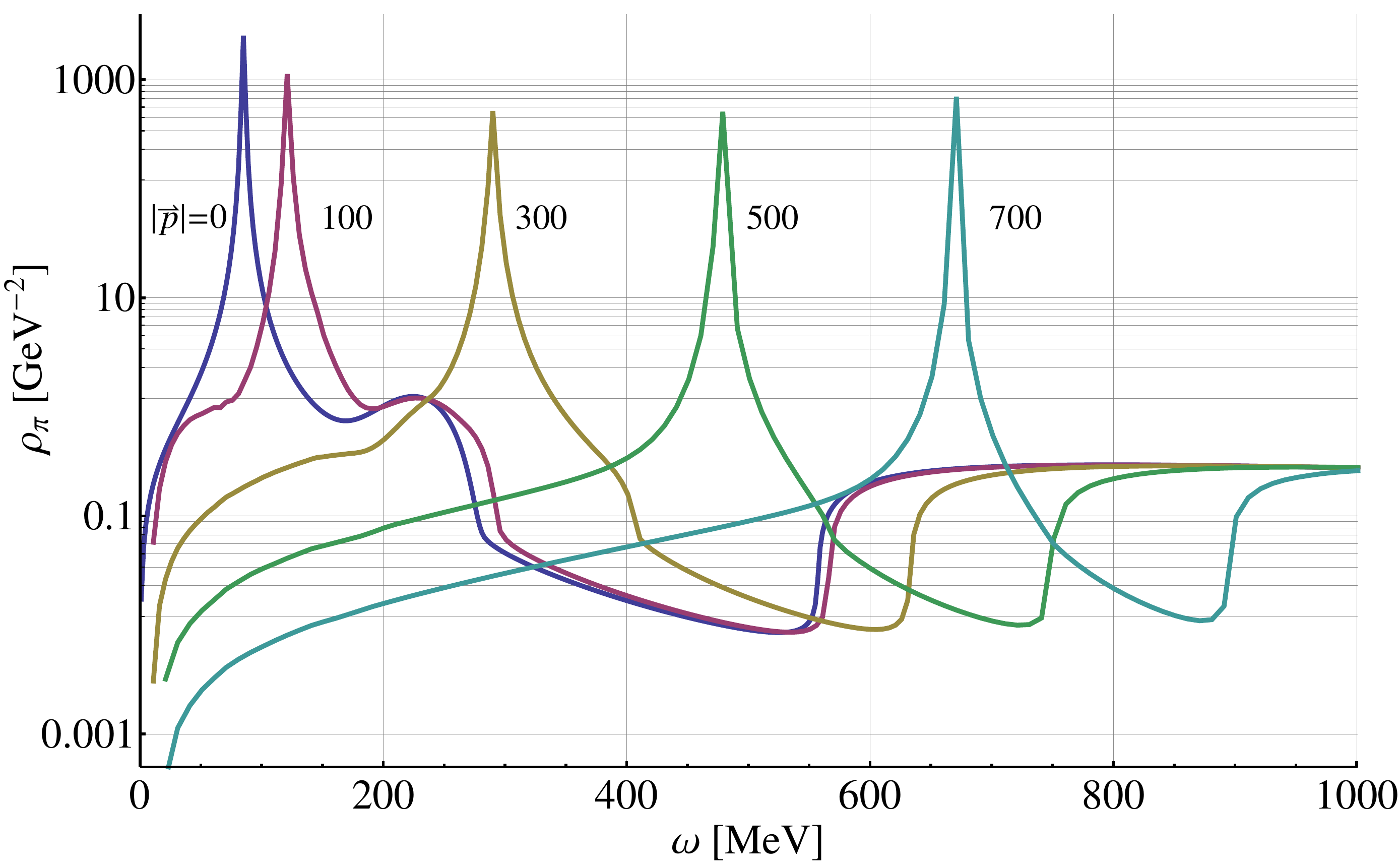}
\caption{(color online) The sigma (left) and pion (right) spectral function, $\rho_\sigma (\omega, \vec{p})$ and $\rho_\pi (\omega, \vec{p})$, are shown versus external energy~$\omega$ at $T=100$~MeV and $\mu=0$~MeV for different external momenta $|\vec{p}|$: 0~MeV~(blue), 100~MeV~(magenta), 300~MeV~(ochre), 500~MeV~(green), 700~MeV~(turquoise).}
\label{fig:spectral_results}
\end{figure*}

The sigma spectral function, $\rho_\sigma (\omega, \vec{p})$, is shown in Fig.~\ref{fig:spectral_results} at a temperature of $T=100$~MeV and for different external spatial momenta~$\vec{p}$. For zero spatial momentum the sigma spectral function still closely resembles its vacuum shape, cf.\ Fig.~\ref{fig:lorentz} in App.~\ref{app:lorentz}, which is dominated by the decay channels into two pions and into a quark anti-quark pair. Using Eqs. (\ref{eq:sigma_pion_pion_decay}) and (\ref{eq:sigma_quark_quark_decay}), the thresholds for these processes are found to be $\omega \geq 2m_\pi\approx 284$~MeV and $\omega \geq 2m_\psi\approx 558$~MeV while the decay into two sigma mesons occurs for $\omega \geq 2m_\sigma\approx 850$~MeV and has only little effect on the spectral function, cf.\ Fig.~\ref{fig:spectral_results}. 

For finite external spatial momenta we observe two main effects. On one hand, the time-like portion of the spectral function, $\omega\geq |\vec{p}|$, is boosted to higher energies, as expected from Lorentz covariance, cf.\ App.~\ref{app:lorentz}, and from the kinematic constraints in Eqs.~(\ref{eq:sigma_sigma_sigma_decay})-(\ref{eq:sigma_quark_quark_decay}).\footnote{We also note that a small peak forms near the two-pion threshold at larger $|\vec{p|}$, cf.\ the discussion in App.~\ref{app:lorentz} on the effects of the violation of Lorentz invariance.} On the other hand, the space-like processes given by Eq.~(\ref{eq:spacelike_sigma}) give rise to an increase of the spectral function in the regime $0\leq \omega \leq |\vec{p}|$, where the largest contribution is found to stem from the absorption or emission of a sigma excitation by a quark or anti-quark. With increasing momentum $|\vec{p}|$ this new space-like regime is broadened and flattened due to the fact that the same range of loop momenta gives now rise to space-like processes in a larger $\omega$-range, cf.\ the explicit example given for the pion spectral function below. 

The pion spectral function, $\rho_\pi (\omega, \vec{p})$, as shown on the right-hand side of Fig.~\ref{fig:spectral_results}, exhibits a strong peak at $\omega\approx 100$~MeV for zero external spatial momentum which is the pion pole contribution and which is due to the zero-crossing of the real part of the corresponding 2-point function. Apart from that, the spectral function is most strongly affected by the decay into two quarks, cf.\ Eq.~(\ref{eq:pion_quark_quark_decay}), and the thermal process where one pion is absorbed from the heat bath and a sigma meson is produced, cf.\ Eq.~(\ref{eq:pion_pion_sigma_decay_T}), which is possible for $\omega \leq m_\sigma - m_\pi\approx 283$~MeV, cf.\ Fig.~\ref{fig:spectral_results}.

For finite external spatial momenta we again observe the expected Lorentz-boosting of the spectral function. The effect of the space-like processes is not as pronounced as for the sigma spectral function, which can be explained as follows. When comparing Eqs.~(\ref{eq:pion_pion_sigma_decay_T})-(\ref{eq:pion_sigma_pion_decay_T}) and the corresponding expressions in Eq.~(\ref{eq:spacelike_pion}) we note that they describe essentially the same processes and that it depends on the loop momentum whether the process classifies as time-like or space-like. For example, if we take the loop momentum to be zero, $\vec{q}=0$, energy conservation yields that ${\omega=\sqrt{m_\sigma^2+\vec{p}^{\,2}}-m_\pi}$ for the process ${\pi^* + \pi \rightarrow \sigma}$, which for $T=100$~MeV and $|\vec{p}|=100$~MeV leads to a time-like energy of $\omega\approx 295$~MeV. On the other hand, for a loop momentum of, e.g., $\vec{q}=-5\vec{p}$, one gets $\omega\approx 64$~MeV, clearly a space-like configuration. This explicit example illustrates the fact that the integration over the loop momentum can give rise to space-like as well as time-like contributions corresponding to the same process and that we therefore observe only one threshold associated with the process ${\pi^* + \pi \rightarrow \sigma}$, with the upper bound given in Eq.~(\ref{eq:pion_pion_sigma_decay_T}). The process ${\pi^* + \sigma \rightarrow \pi}$, on the other hand, only gives rise to a space-like contribution since $m_\pi-m_\sigma<0$ at $T=100$~MeV. Finally, we note that the effect of the time-like process ${\pi^* + \pi \rightarrow \sigma}$ decreases with higher momentum $|\vec{p}|$, cf.\ Fig.~\ref{fig:spectral_results}. This is in agreement with the expectation that a particle propagating through a heat bath at a momentum significantly larger than average thermal fluctuations, i.e.\ $|\vec{p}|\gg T$, should experience no significant modifications from the medium.

\begin{figure*}[t]
\includegraphics[width=\columnwidth]{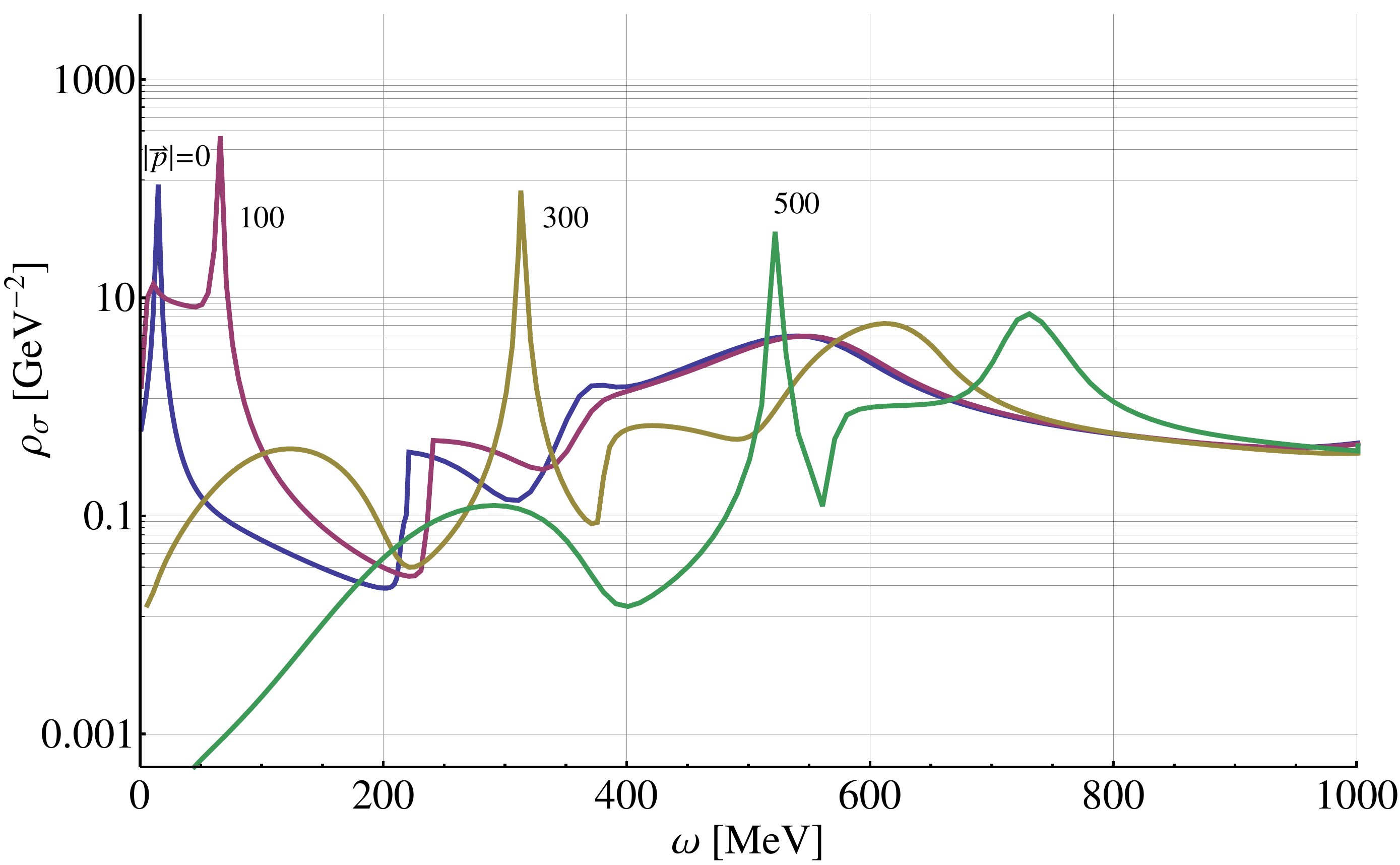}\hspace{3mm}
\includegraphics[width=\columnwidth]{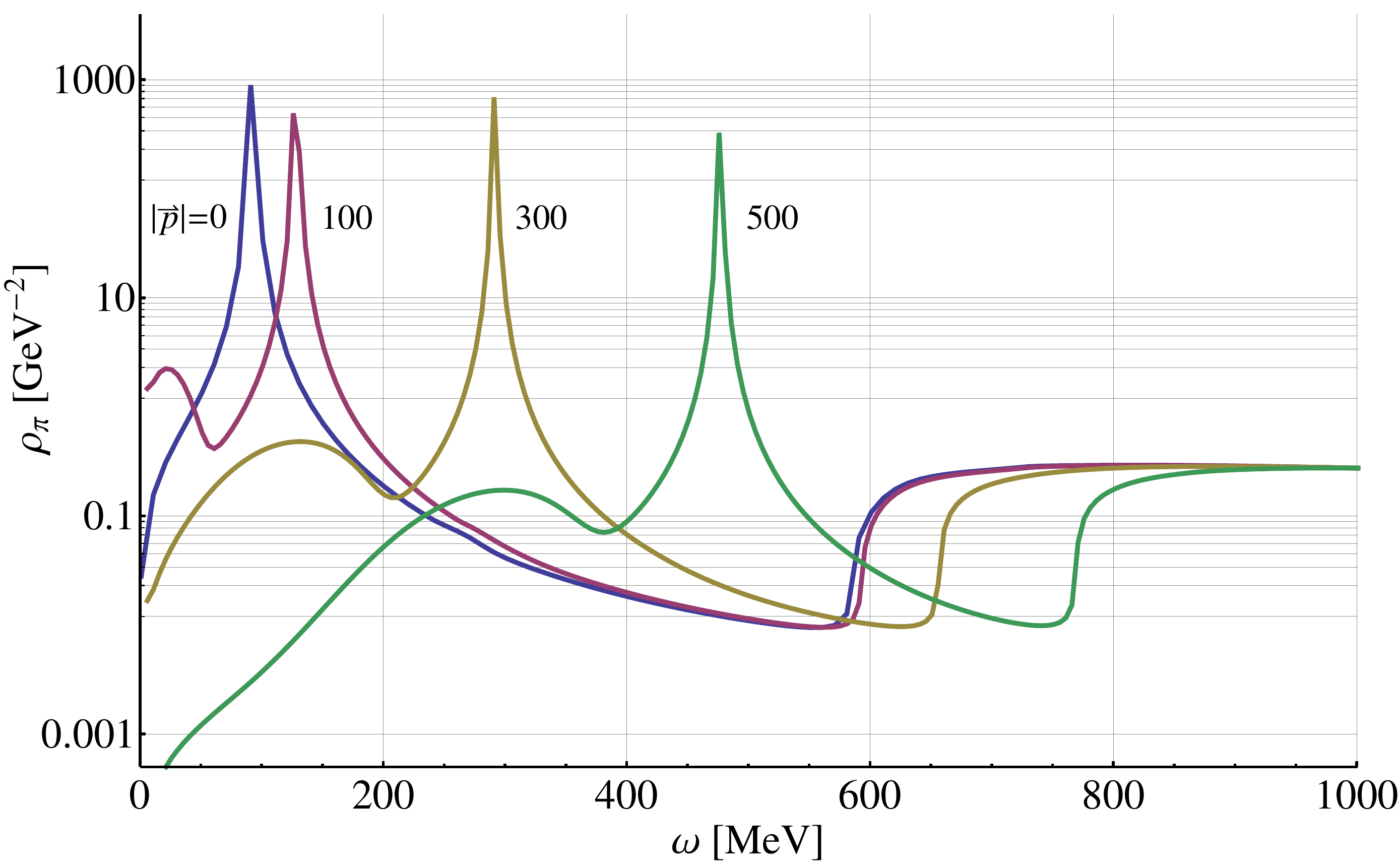}
\caption{(color online) The sigma and pion spectral function, $\rho_\sigma (\omega, \vec{p})$ and $\rho_\pi (\omega, \vec{p})$, are shown versus external energy $\omega$ at $T=10$~MeV and $\mu=292.97$~MeV for different external momenta $|\vec{p}|$: 0~MeV~(blue), 100~MeV~(magenta), 300~MeV~(ochre), 500~MeV~(green).}
\label{fig:spectral_results_CEP}
\end{figure*}

We now discuss the spatial-momentum dependence of the sigma and pion spectral function near the CEP, cf.\ Fig.~\ref{fig:spectral_results_CEP} which shows $\rho_\sigma (\omega, \vec{p})$ and $\rho_\pi (\omega, \vec{p})$ at a chemical potential of $\mu=292.97$~MeV and a temperature of $T=10$~MeV for different external spatial momenta.
We note that the quark-meson model is not intended to provide a quantitatively reliable description of the QCD phase diagram at such large chemical potentials since explicit baryonic (and gluonic) degrees of freedom are not included. In particular the mixing of the sigma fluctuations with those of the baryon density in the nuclear matter region, and possibly also with electric glueball correlations \cite{Hatta2004}, might lead to significant changes, e.g., in the location of the CEP and the phase structure at higher densities. However, we expect that the qualitative effects discussed in the following, such as the sigma spectral function developing a pole at zero energy near the CEP, are to a large extent universal and not qualitatively affected by baryonic or gluonic fluctuations.

At $|\vec{p}|=0$ we find that the sigma spectral function exhibits a peak at very small energies, which is due to a zero-crossing of the real part of the sigma 2-point function, and which signals a stable particle pole at mass given by the peak position. This is in agreement with the expectation that the radial $\sigma$-field becomes massless at the CEP, i.e.\ at a second-order phase transition. At higher energies, the first threshold that appears in $\rho_\sigma$ is due to the decay into two sigma mesons, cf.\ Eq.~(\ref{eq:sigma_sigma_sigma_decay}) with $m_\sigma \approx 106$~MeV. Since the thresholds are determined by the curvature masses of the quasi-particles, cf.\ Sec.~\ref{sec:processes}, we note that the curvature and pole mass of the sigma meson do not agree in this case. This is expected in our present truncation because the curvature mass changes more rapidly in the vicinity of the CEP without wavefunction renormalization.  They both approach zero at the CEP, however, as they must, cf.\ also \cite{Tripolt2014}. 

Beyond this threshold we encounter some numerical problems for small values of $\epsilon$ which we circumvent by using $\epsilon=20$~MeV in an energy range of $\Delta\omega\approx 200$~MeV beyond the sigma-sigma threshold for $|\vec{p}|=0$~MeV and $|\vec{p}|=100$~MeV. This method is justified in App.~\ref{app:epsilon}, where it is shown that the spectral functions in our calculation are rather insensitive to $\epsilon$ in energy regimes that allow for quasi-particle processes. We have also checked explicitly that the sigma spectral function shown in Fig.~\ref{fig:spectral_results_CEP} does not depend on $\epsilon$ at higher energies, where the aforementioned numerical problems do not arise.

For finite external spatial momenta the sigma spectral function is affected by the same space-like processes as described before, which are now, however, suppressed by the low temperature. For small momenta $\vec{p}$, these space-like processes affect the shape of the peak in the sigma spectral function, cf.\ Fig.~\ref{fig:spectral_results_CEP}. This interference of the space-like part and the peak at time-like energies is, however, only possible due to the finite value for $\epsilon$. In the limit $\epsilon\rightarrow 0$, the peak will turn into a delta function, leaving the space-like part clearly separated, cf.\ the discussion in App.~\ref{app:epsilon}. At higher spatial momenta we observe that these two effects separate anyhow as the spectral function is boosted to higher energies. 

Finally, we discuss the momentum dependence of the pion spectral function near the CEP, as shown on the right panel of Fig.~\ref{fig:spectral_results_CEP}. At $\vec{p}=0$ the pion spectral functions closely resembles its vacuum shape, cf.\ App.~\ref{app:epsilon}. At higher momenta we observe similar modifications in the space-like regime as for the sigma spectral function. This is due to the fact that the thresholds for the time-like processes ${\pi^* + \pi \rightarrow \sigma}$ and ${\pi^* + \sigma \rightarrow \pi}$, cf.\ Eqs.~(\ref{eq:pion_pion_sigma_decay_T})-(\ref{eq:pion_sigma_pion_decay_T}), are now located at very small time-like energies since ${m_\sigma \approx 106}$~MeV and ${m_\pi\approx 139}$~MeV in this critical regime. Therefore, the integration over the loop momentum now gives a large contribution for space-like energies, a situation similar as to the case of the sigma spectral function, where all thermal processes are anyhow limited to the space-like region. Besides these modifications at space-like energies, we observe that the pion spectral function is boosted to higher energies with increasing spatial momentum, as expected.

\section{Summary and Outlook}\label{sec:summary}

In this work we have studied the spatial-momentum dependence of mesonic spectral functions by using the quark-meson model within a non-perturbative FRG approach. The necessary analytic continuation from imaginary to real energies is achieved on the level of the flow equations without the need for numerical reconstruction schemes. Moreover, this method preserves the breaking pattern of chiral symmetry, is thermodynamic consistent and can in principle be extended to full QCD. 

Results were presented for the sigma and pion spectral function at finite temperatures and densities, in particular near the CEP in the phase diagram of the quark-meson model. It was found that the time-like region of the spectral functions is boosted towards higher energies while the regime with energies smaller than the spatial momentum is affected by space-like processes. The influence of space-like processes is found to be weaker for the pion spectral function, which can be explained by the fact that there is only one process that occurs exclusively for space-like momentum configurations, i.e.\ the absorption or emission of a pion excitation by a quark or an anti-quark.

Near the CEP, the sigma spectral function exhibits a peak at very small energies and zero spatial momentum, signaling a stable particle with an almost vanishing pole mass near this second order phase transition. For finite external spatial momenta, this peak at low energies almost mixes with the space-like processes until these two effects separate at larger momenta. The pion spectral function, on the other hand, remains basically unchanged as compared to its vacuum shape at zero spatial momentum while it experiences similar modifications as the sigma spectral functions from space-like processes at higher momenta.

We believe this work to be a worthwhile first step towards a more quantitative calculation of the momentum dependence of real-time quantities like spectral functions which will allow, e.g., for the calculation of transport coefficients like the shear viscosity at various temperatures and densities, in particular within critical regions of the phase diagram.

\acknowledgments 
We thank D.~Rischke, J.~Pawlowski, N.~Strodthoff, B.-J.~Schaefer and L. McLerran for valuable discussions. This work was supported by the Helmholtz International Center for FAIR within the LOEWE initiative of the State of Hesse. \mbox{R.-A.T.} is furthermore supported by the Helmholtz Research School for Quark Matter Studies, H-QM. Our numerical calculations were carried out on the Lichtenberg-Hochleistungsrechner in Darmstadt and on the LOEWE-CSC in Frankfurt.

\appendix

\begin{figure*}[t]
\includegraphics[width=\columnwidth]{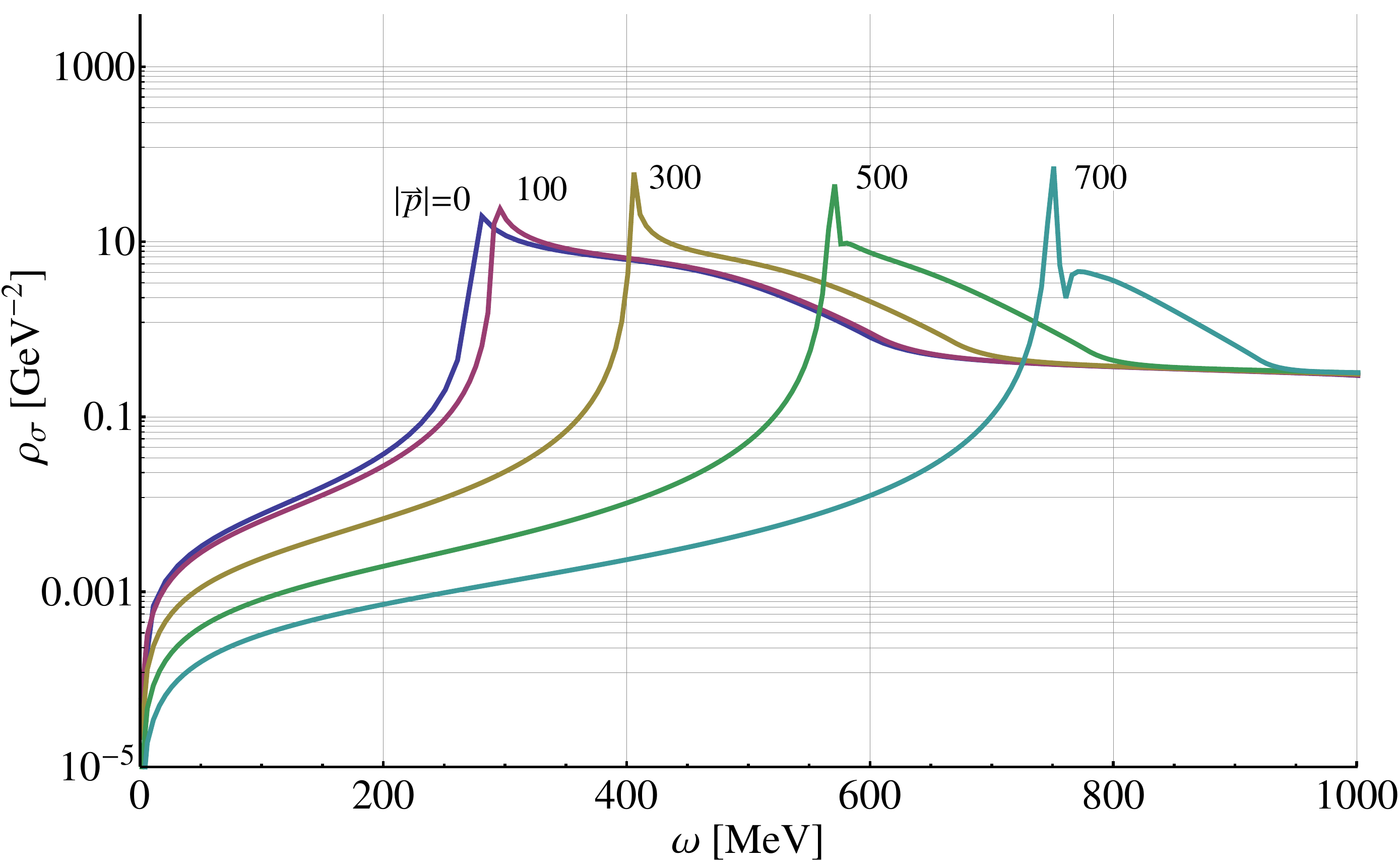}\hspace{3mm}
\includegraphics[width=\columnwidth]{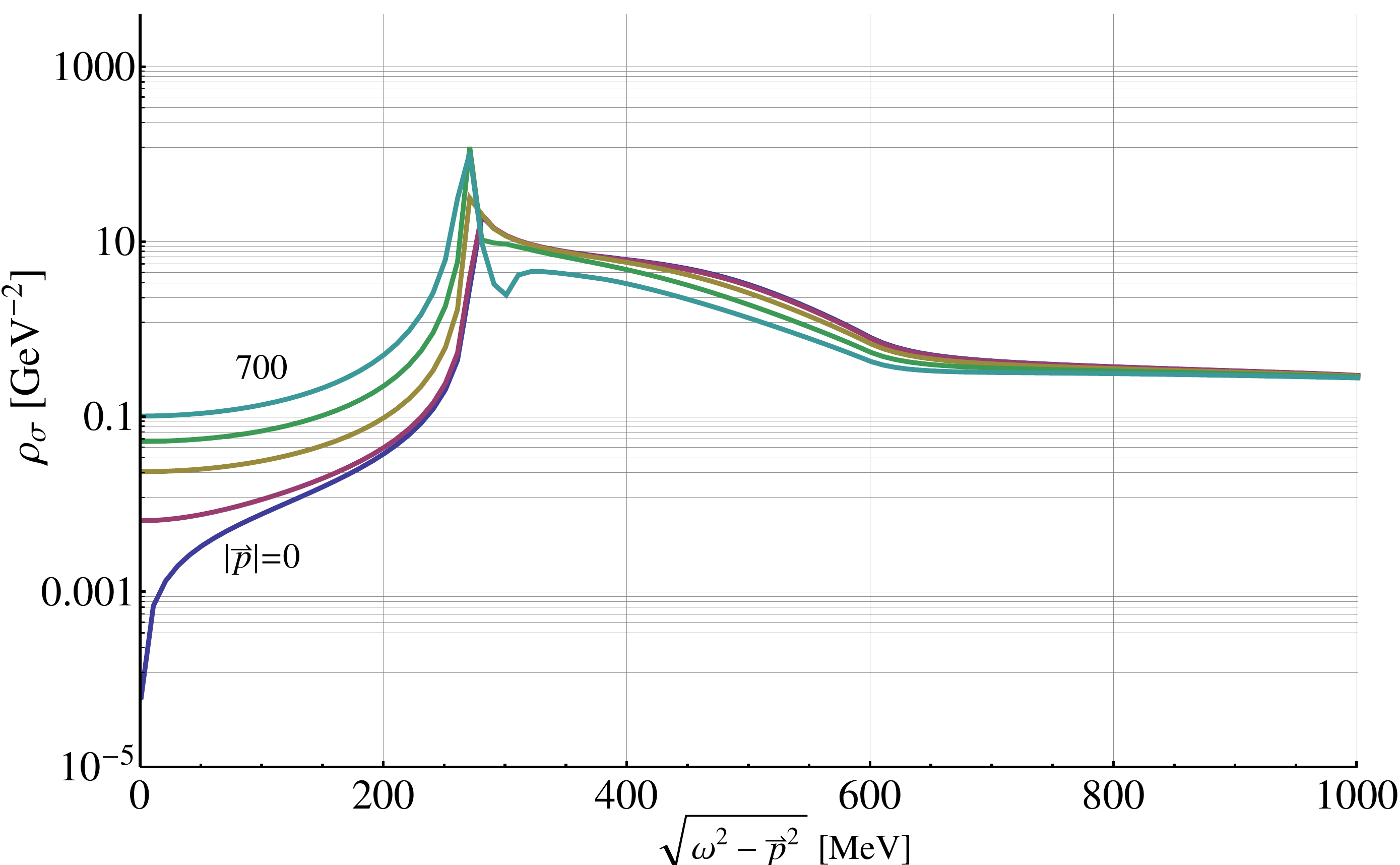}
\caption{(color online) The sigma spectral function, $\rho_\sigma (\omega, \vec{p})$, is shown as a function of external energy $\omega$ (left) and as a function of the invariant $\sqrt{\omega^2-\vec{p}^{\,2}}$ (right) at $T=0$~MeV and $\mu=0$~MeV for different external momenta $|\vec{p}|$: 0~MeV~(blue), 100~MeV~(magenta), 300~MeV~(ochre), 500~MeV~(green), 700~MeV~(turquoise).}
\label{fig:lorentz}
\end{figure*}

\section{Lorentz Invariance}\label{app:lorentz}
In this section we study the extent of the breaking of Lorentz or O(4) invariance induced by the use of dimensionally reduced regulator functions, cf.\ Eqs.~(\ref{eq:3dregulators})-(\ref{eq:3dregulators2}). Such three-dimensional regulator functions only regulate the spatial momentum but allow for arbitrarily large energy transfers which may give rise to a violation of Lorentz invariance, cf.\ also \cite{Kamikado2014}. In order to study only the effect due to the 3d regulator functions and not from, e.g. the heat bath, we will focus on $T=0$ and $\mu=0$ in the following.

The sigma spectral function is shown for different external spatial momenta in Fig.~\ref{fig:lorentz}. We note that, when plotted as a function of external energy, the threshold for the two-pion decay channel, cf.\ Eq.~(\ref{eq:sigma_pion_pion_decay}), is boosted to higher energies. From Lorentz invariance we would expect the shift in energy to be 
\begin{equation}
\Delta\omega= \sqrt{\omega_0^2+\vec{p}^{\,2}}-\omega_0,
\end{equation}
where $\omega_0$ denotes the location of the peak or threshold at $\vec{p}=0$. Using $\omega_0\approx 270$~MeV, we find the expected boost to be $\Delta\omega\approx 480$~MeV for a momentum of $|\vec{p}|=700$~MeV, which agrees with the shift observed in Fig.~\ref{fig:lorentz}. 

However, although the extent of the Lorentz boost is found to be correct, the shape of the sigma spectral function does change slightly when increasing the spatial momentum, i.e.\ it develops a peak near the two-pion threshold. This effect has been found to be due to a ($\omega$-independent) shift of the real part of the 2-point function to higher values, causing its zero crossing to be shifted to smaller energies. This effect is also found for the pion spectral function and could be corrected for by shifting the real part of the 2-point function back to its original value by hand, in order to recover Lorentz invariance, which is, however, not done in the present work.

When plotting the sigma spectral function vs. the invariant $\sqrt{\omega^2-\vec{p}^{\,2}}$, as done on the right-hand side of Fig.~\ref{fig:lorentz}, one would expect, given Lorentz invariance, that all curves lie on top of each other. With increasing spatial momentum we do, however, observe some deviations from the curve for $\vec{p}=0$. Apart from the already discussed formation of a peak, we observe that the spectral function exhibits larger values in the low-energy regime with increasing spatial momenta. This is due to the employed UV shape of the 2-point functions, cf.\ Eqs.~(\ref{eq:UV_sigma})-(\ref{eq:UV_pion}), which in clearly not Lorentz-invariant for $\epsilon > 0$, cf.\ also \cite{Kamikado2014}. In fact, this effect is expected to disappear in the limit $\epsilon\rightarrow 0$, cf.\ App.~\ref{app:epsilon}.

We conclude that the extent of the breaking of Lorentz symmetry induced by the use of three-dimensional regulator functions is small and that its effects are well understood, allowing for a robust study of the momentum dependence of spectral functions.

\begin{figure*}[t]
\includegraphics[width=\columnwidth]{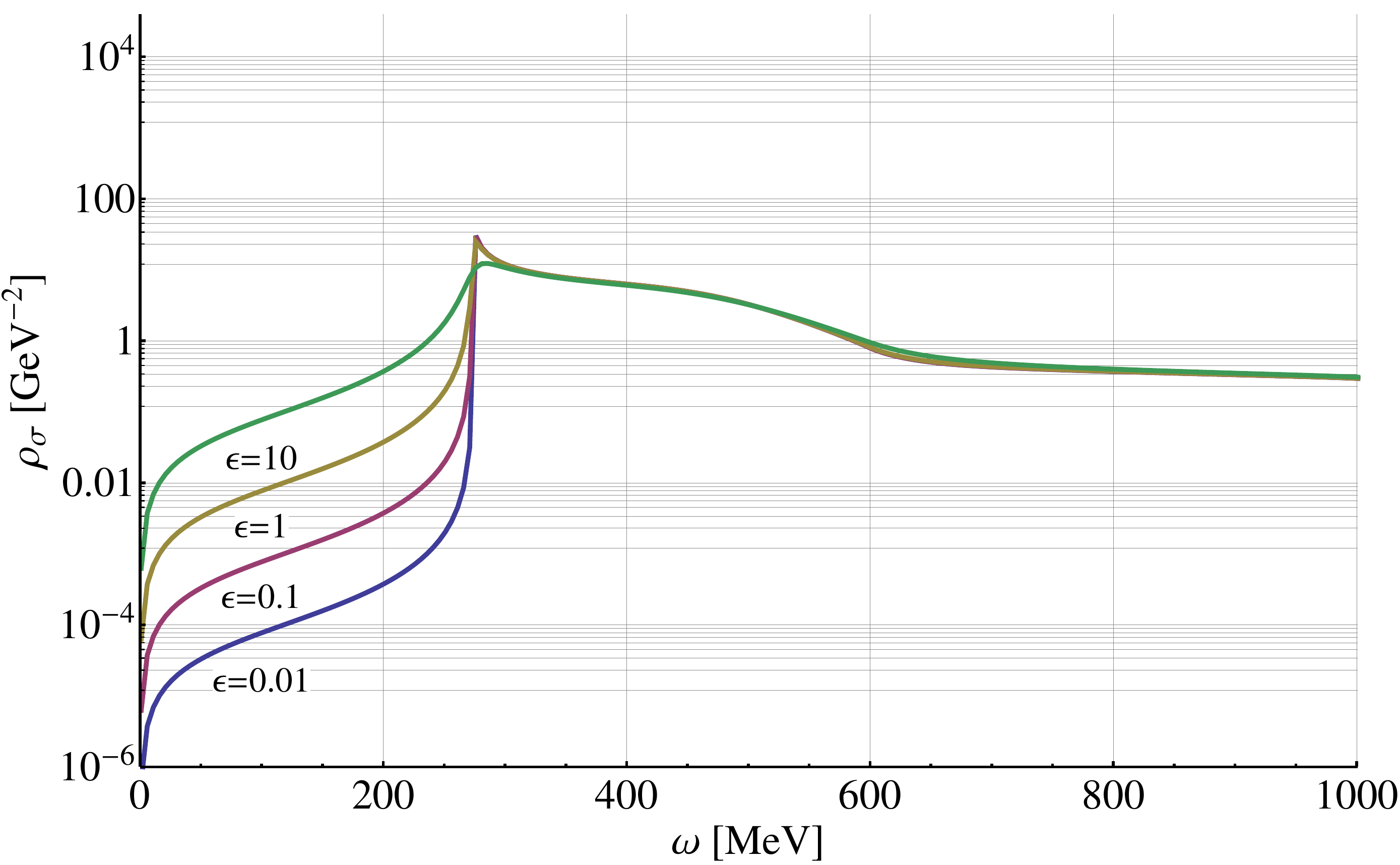}\hspace{3mm}
\includegraphics[width=\columnwidth]{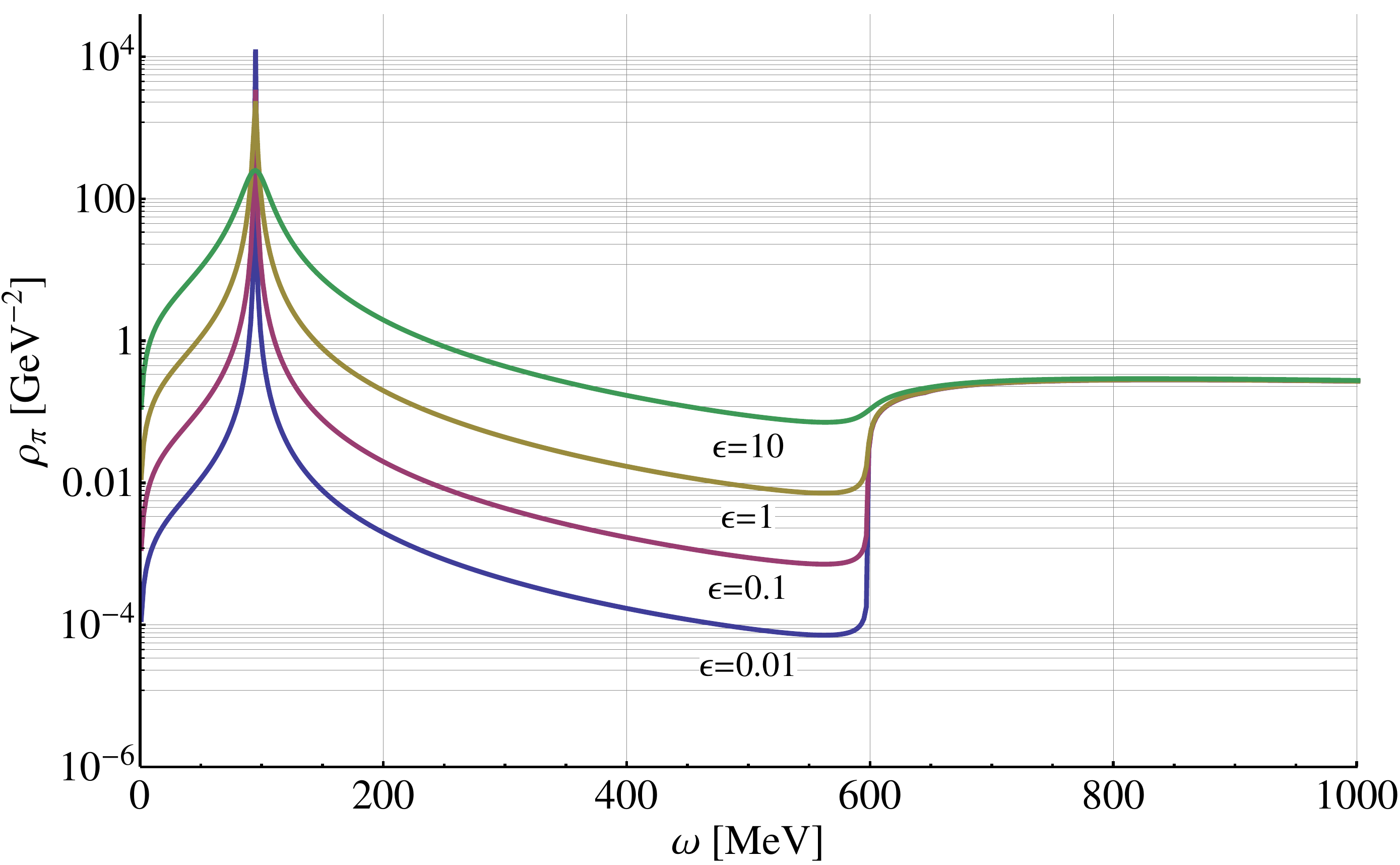}
\caption{(color online) The sigma (left) and pion (right) spectral function, $\rho_\sigma (\omega, \vec{p})$ and $\rho_\pi (\omega, \vec{p})$, are shown versus external energy $\omega$ at $T=0$~MeV and $\mu=0$~MeV for different values of the parameter $\epsilon$: 0.01~MeV~(blue), 0.1~MeV~(magenta), 1~MeV~(ochre) and 10~MeV~(green).}
\label{fig:epsilon}
\end{figure*}

\section{Parameter dependence}\label{app:epsilon}
In this section we study the dependence of our results on the parameter $\epsilon$, as introduced in the definition for the retarded 2-point function, cf.\ Eq.~(\ref{eq:continuation2}), which is reproduced here for convenience,
\begin{equation}
\label{eq:continuation2_copy}
\Gamma^{(2),R}(\omega,\vec p)=-\lim_{\epsilon\to 0} \Gamma^{(2),E}(p_0=-\I(\omega+\I\epsilon), \vec p).
\end{equation}
In principle, the retarded 2-point functions are only obtained in the limit $\epsilon\rightarrow 0$, which is, however, not possible to achieve exactly in a numerical calculation. Moreover, the imaginary part of the 2-point functions vanishes for $\epsilon=0$, cf.\ Eqs.~(\ref{eq:J_B_pi})-(\ref{eq:delta_J_tilde}) with $p_0=-\I\omega$, and therefore also the spectral functions will be zero, cf.\ Eq.~(\ref{eq:spectralfunction}). We do, however, find, that our results depend on a clear way on $\epsilon$, as described in the following, and that our results are indeed meaningful, even for a finite value of $\epsilon$.

The sigma and pion spectral function are shown in Fig.~\ref{fig:epsilon} for vacuum conditions and different values of $\epsilon$. We note that, at energies for which no decay channels are available, the value of the spectral functions in general decreases with smaller $\epsilon$ and is expected to vanish in the limit $\epsilon \rightarrow 0$. When the spectral function, however, exhibits a peak that is to be associated with a stable particle, this peak is expected to turn into a Dirac delta function in the limit $\epsilon \rightarrow 0$. Such a behavior is indeed observed for the pion spectral function in Fig.~\ref{fig:epsilon}.

On the other hand, at energies that allow for quasi-particle processes to take place, cf.\ Eqs.~(\ref{eq:sigma_sigma_sigma_decay})-(\ref{eq:spacelike_pion}), we observe that the spectral functions do not depend on $\epsilon$, cf.\ Fig.~\ref{fig:epsilon}. Our results are therefore expected to reproduce the correct retarded result whenever there are quasi-particle processes available, while for energy regimes where there are no processes possible, the spectral functions will be either zero or reduce to a delta function in the limit~$\epsilon\rightarrow 0$.

As described in Sec.~\ref{sec:results_p}, we utilize this finding by employing a larger value for $\epsilon$ to calculate the sigma spectral function near the critical endpoint at certain energies, in order to circumvent numerical problems that arise for smaller $\epsilon$.

\section{Threshold functions}\label{app:thresholds}
In this section we list explicit expressions for the threshold and loop functions appearing in the flow equations for the effective potential, Eq.~(\ref{eq:flow_pot}), and the mesonic two-point functions, Eqs.~(\ref{eq:Gamma2sigmaB})-(\ref{eq:Gamma2pionF}), using the 3d regulator functions from Eqs.~(\ref{eq:3dregulators})-(\ref{eq:3dregulators2}).

The bosonic threshold functions are given by
\begin{eqnarray}
I_{k,\alpha}^{(1)} &=&
\frac{k^4}{6\pi^2}
\frac{1+2n_B(E_{k,\alpha})}{E_{k,\alpha}}
\, , \\
I_{k,\alpha}^{(2)} &=&
\frac{k^4}{6\pi^2}
\left(
\frac{1+2n_B(E_{k,\alpha})}{2E_{k,\alpha}^3}-
\frac{n_B'(E_{k,\alpha})}{E_{k,\alpha}^2}
\right),
\end{eqnarray}
with $\alpha \in \{\pi,\sigma\}$, while the fermionic threshold function can be written as
\begin{eqnarray}
I_{k,\psi}^{(1)} &=&
\frac{k^4}{3\pi^2}
\frac{1-n_F(E_{k,\psi}-\mu)-n_F(E_{k,\psi}+\mu)}{E_{k,\psi}},
\end{eqnarray}
with the bosonic and fermionic occupation numbers,
\begin{equation}
\label{eq:n_B_F} 
n_B(E)\equiv\frac{1}{e^{E/T}-1},\quad
n_F(E)\equiv\frac{1}{e^{E/T}+1}
\, .
\end{equation}
The effective quasi-particle energies read
\begin{equation}
\label{eq:energies}
E_{k,\alpha}\equiv\sqrt{k^2+m_{k,\alpha}^2}, \qquad \alpha \in
\{\pi,\sigma,\psi\}\,, 
\end{equation}
where the effective meson masses and the quark mass are given by
\begin{equation}
\label{eq:masses}
m_{k,\pi}^2=2U_k',\quad 
m_{k,\sigma}^2=2U_k'+4 U_k''\phi^2,\quad
m_{\psi}^2=h^2\phi^2\,,
\end{equation}
where primes denote derivatives with respect to $\phi^2$, with $\phi \equiv(\sigma,\pi_1,\pi_2,\pi_3)$. The mesonic vertex functions defined in Eqs.~(\ref{eq:meson_vertex_3})-(\ref{eq:meson_vertex_4}) read explicitly
\begin{eqnarray}
\Gamma_{k,\sigma\sigma\sigma}^{(0,3)}&=&12U_k''\phi+8U_k^{(3)}\phi^3, \\
\Gamma_{k,\sigma\pi\pi}^{(0,3)}&=&4U_k''\phi, \\
\Gamma_{k,\sigma\sigma\sigma\sigma}^{(0,4)}&=&12U_k''+48U_k^{(3)}\phi^2+16U_k^{(4)}\phi^4,
\\
\Gamma_{k,\pi\pi\tilde{\pi}\tilde{\pi}}^{(0,4)}&=&4U_k'',\\
\Gamma_{k,\pi\pi\pi\pi}^{(0,4)}&=&12U_k'',\\
\Gamma_{k,\sigma\sigma\pi\pi}^{(0,4)}&=&4U_k''+8U_k^{(3)}\phi^2\,,
\end{eqnarray}
with $\pi,\tilde{\pi}\in \{\pi_1,\pi_2,\pi_3\}$ and $\pi\neq\tilde{\pi}$.

We now present explicit expressions for the bosonic and fermionic loop functions, $J_{k,\alpha\beta}(p)$ and $J_{k,\bar{\psi}\psi}^{(\alpha)}(p)$, where the first step of our analytic continuation procedure, given by Eq.~(\ref{eq:continuation1}) has already been carried out, i.e.\ the external energy $p_0$ just has to be replaced by 
\begin{equation}
p_0\rightarrow-\I(\omega+\I\epsilon)
\end{equation}
in order to obtain the retarded expressions. We note that the bosonic and fermionic loop functions crucially depend on the configuration of the loop momentum~$\vec{q}$, the external momentum~$\vec{p}$ and the RG scale~$k$ due to the appearance of a theta function of the form
\begin{equation}
\Theta(k^2-(\vec{q}+\vec{p})^2)=\begin{cases}1, &\quad |\vec{q}+\vec{p}|\leq k\\
0, &\quad |\vec{q}+\vec{p}|>k \end{cases}
\end{equation}
which arises from the 3d regulator functions, cf.\ Eqs.~(\ref{eq:3dregulators})-(\ref{eq:3dregulators2}), and subsequently also appears in the general expression for the loop functions, cf.\ Eqs.~(\ref{eq:J_def})-(\ref{eq:J_F_def}). For momentum configurations with $|\vec{q}+\vec{p}|\leq k$, the bosonic loop functions, $J_{k,\alpha\beta}(p)$, are then given by Eq.~(\ref{eq:J_B}),\footnote{We note that the RG-scale index $k$ of the energies $E_{k,\alpha}$ has been omitted in Eqs.~(\ref{eq:J_B_pi})-(\ref{eq:delta_J_tilde}) for simplicity.} where $\alpha\in \{ \sigma,\pi\}$ always represents the particle associated with the ``upper" leg of the loops, i.e.\ the one with the regulator insertion, and $\beta\in \{\sigma,\pi\}$ always represents the particle associated with the ``lower" leg of the loops, i.e.\ the one without regulator insertion, cf.\ Fig.~\ref{fig:flow_Gamma2}. For momentum configurations with $|\vec{q}+\vec{p}|>k$, the bosonic loop functions are denoted as $\tilde{J}_{k,\alpha\beta}(p)$ and are obtained by carrying out the following substitution in Eq.~(\ref{eq:J_B}),
\begin{equation}
\tilde{J}_{k,\alpha\beta}\equiv J_{k,\alpha\beta}\left(E_\beta\rightarrow\tilde{E}_\beta\equiv\sqrt{E_\beta^2-k^2+\vec{q}^{\,2}}\,\right).
\end{equation}

Similarly, the fermionic loop function for the pion 2-point function $J_{k,\bar{\psi}\psi}^{(\pi)}(p)$ is, for momentum configurations with $|\vec{q}+\vec{p}|\leq k$, given by Eq.~(\ref{eq:J_pi}), where 
\begin{equation}
\cos \varphi \equiv \frac{\vec{q}\cdot(\vec{q}+\vec{p})}{|\vec{q}||\vec{q}+\vec{p}|}.
\end{equation}
The corresponding expression for the sigma 2-point function can be written as
\begin{equation}
J^{(\sigma)}_{k,\bar{\psi}\psi}(p)\equiv J^{(\pi)}_{k,\bar{\psi}\psi}(p)+\Delta J_{k,\bar{\psi}\psi}(p),
\end{equation}
where ${\Delta J_{k,\bar{\psi}\psi}(p)}$ is given by Eq.~(\ref{eq:delta_J}). For momentum configurations with $|\vec{q}+\vec{p}|>k$, the fermionic loop function for the pion 2-point function, $\tilde{J}^{(\pi)}_{k,\bar{\psi}\psi}(p)$, is given by Eq.~(\ref{eq:J_pi_tilde}) while 
\begin{equation}
{\Delta \tilde{J}_{k,\bar{\psi}\psi}(p) \equiv \tilde{J}^{(\sigma)}_{k,\bar{\psi}\psi}(p)-\tilde{J}^{(\pi)}_{k,\bar{\psi}\psi}(p)}
\end{equation}
is given by Eq.~(\ref{eq:delta_J_tilde}), where
\begin{equation}
\tilde{E}_\psi\equiv\sqrt{E_\psi^2-k^2+\vec{q}^{\,2}}.
\end{equation}
We note that the different contributions to the bosonic and fermionic loop functions can be readily interpreted in terms of particle creation and annihilation processes. 
As a simple example we discuss the bosonic loop function $J_{k,\pi\pi}(p)$ which contributes to the sigma 2-point function. The explicit expression of $J_{k,\pi\pi}(p)$ can be worked out by using Eq.~(\ref{eq:J_B}) and is presented in Eq.~(\ref{eq:J_B_pi}) for convenience. Therein, the term ${(1+n_B(E_\pi))(1+n_B(E_\pi))}$ represents the statistical weight factor for the direct process ${\sigma^*\rightarrow\pi+\pi}$, whereas the term ${n_B(E_\pi)n_B(E_\pi)}$ represents the statistical weight factor for the inverse process ${\pi+\pi\rightarrow\sigma^*}$, which is only possible at finite temperature since real particles from the heat bath are needed. Similarly, the term $n_B(E_\pi)(1+n_B(E_\pi))$ represents the weight factor for the processes ${\sigma^*+\pi\rightarrow\pi}$ and ${\pi\rightarrow\sigma^*+\pi}$ which are both only possible at finite temperature. We note that, when calculating the imaginary part of the loop functions explicitly and taking the limit $\epsilon\rightarrow 0$, the fractional expressions collapse to Dirac $\delta$-functions like $\delta (\omega-2E_\pi)$, see e.g. \cite{Das:1997gg} for a more detailed discussion.

For the fermionic loop functions the statistical factors have the form $(1-n_F(E_\psi))$, representing the fact that there are real fermions in the heat bath, which gives rise to Pauli blocking, i.e.\ the available density of states for the decay products is suppressed.

\begin{widetext}
\begin{equation}
\label{eq:J_B_pi}
		\begin{split}
			J_{k,\pi\pi}(p) =\frac{k}{2}
			\Biggl(&
			((1+n_B(E_\pi))(1+n_B(E_\pi))-n_B(E_\pi)n_B(E_\pi))
			\frac{12E_\pi^2+p_0^2}{E_\pi^3(4E_\pi^2+p_0^2)^2}\\
			&+n_B(E_\pi)(1+n_B(E_\pi))
			\frac{1}{E_\pi^2(E_\pi^2-(E_\pi-\I p_0)^2)T}
			+
			n_B(E_\pi)(1+n_B(E_\pi))
			\frac{1}{E_\pi^2(E_\pi^2-(E_\pi+\I p_0)^2)T}
			\Biggr)
		\end{split}
\end{equation}

\begin{equation}
\label{eq:J_B}
		\begin{split}
			%%%%%%%%%%%%%%%%%%%%%%%%%%%%%%%%%%%%%%%%%%%%%%%%%%
			J_{k,\alpha\beta}(p) =\frac{k}{2}
			\Biggl( &
			(1+n_B(E_\alpha))\frac{
			E_\alpha^2+E_\beta^2-(2E_\alpha+\I p_0)^2}
			{E_\alpha^3(E_\beta^2-(E_\alpha+\I p_0)^2)^2}
			+n_B(E_\alpha)\frac{
			E_\alpha^2+E_\beta^2-(2E_\alpha-\I p_0)^2}
			{E_\alpha^3(E_\beta^2-(E_\alpha-\I p_0)^2)^2}\\	
			&+(1+n_B(E_\beta))\frac{2}
			{E_\beta(E_\alpha^2-(E_\beta-\I p_0)^2)^2}
			+n_B(E_\beta)\frac{2}
			{E_\beta(E_\alpha^2-(E_\beta+\I p_0)^2)^2}\\
			&+n_B(E_\alpha)(1+n_B(E_\alpha))\frac{1}
			{E_\alpha^2(E_\beta^2-(E_\alpha-\I p_0)^2)T}	
			+n_B(E_\alpha)(1+n_B(E_\alpha))\frac{1}
			{E_\alpha^2(E_\beta^2-(E_\alpha+\I p_0)^2)T}
			\Biggr) \\
			%%%%%%%%%%%%%%%%%%%%%%%%%%%%%%%%%%%%%%%%%%%%%%%%%%
		\end{split}
\end{equation}
\begin{equation}
\label{eq:J_pi}
		\begin{split}
			%%%%%%%%%%%%%%%%%%%%%%%%%%%%%%%%%%%%%%%%%%%%%%%%%%
			J^{(\pi)}_{k,\bar{\psi}\psi}(p) =-2h^2k\Biggl(
			&(1-n_F(E_\psi-\mu))\frac{
			16 E_\psi^4-2 E_\psi^2 (4 E_\psi^2+p_0^2)\cos\varphi
			-k^2(12 E_\psi^2+p_0^2)(1-\cos\varphi)
			}{E_\psi^3(4E_\psi^2+p_0^2)^2}\\
			&-n_F(E_\psi+\mu)\frac{
			8E_\psi^4(2-\cos\varphi)+2E_\psi^2((6k^2-p_0^2)\cos\varphi-6k^2)-k^2p_0^2(1-\cos\varphi)
			}{E_\psi^3(4E_\psi^2+p_0^2)^2}\\
			&-n_F(E_\psi-\mu)(1-n_F(E_\psi-\mu))\frac{
			k^2(1-\cos\varphi)+\I p_0E_\psi
			}{\I p_0E_\psi^2(2E_\psi+\I p_0)T}\\
			&+n_F(E_\psi+\mu)(1-n_F(E_\psi+\mu))\frac{
			k^2(1-\cos\varphi)-\I p_0E_\psi
			}{\I p_0E_\psi^2(2E_\psi-\I p_0)T}
			\Biggr)\\
			%%%%%%%%%%%%%%%%%%%%%%%%%%%%%%%%%%%%%%%%%%%%%%%%%%
		\end{split}
\end{equation}
\begin{equation}
\label{eq:delta_J}
		\begin{split}
			%%%%%%%%%%%%%%%%%%%%%%%%%%%%%%%%%%%%%%%%%%%%%%%%%%
			\Delta J_{k,\bar{\psi}\psi}(p)
			=&\:4m_\psi^2h^2k 
			\Biggl( 
			(1-n_F(E_\psi-\mu))\frac{
			12 E_\psi^2+p_0^2}
			{E_\psi^3(4E_\psi^2+p_0^2)^2}
			-n_F(E_\psi+\mu)\frac{
			12 E_\psi^2+p_0^2}
			{E_\psi^3(4E_\psi^2+p_0^2)^2}\\
			&+n_F(E_\psi-\mu)(1-n_F(E_\psi-\mu))\frac{
			1
			}{\I p_0 E_\psi^2(2E_\psi+\I p_0)T}
			-n_F(E_\psi+\mu)(1-n_F(E_\psi+\mu))\frac{
			1
			}{\I p_0 E_\psi^2(2E_\psi-\I p_0)T}
			\Biggr)\\
			%%%%%%%%%%%%%%%%%%%%%%%%%%%%%%%%%%%%%%%%%%%%%%%%%%
		\end{split}
\end{equation}
\begin{equation}
\label{eq:delta_J_tilde}
		\begin{split}
			%%%%%%%%%%%%%%%%%%%%%%%%%%%%%%%%%%%%%%%%%%%%%%%%%%
			\Delta \tilde{J}_{k,\bar{\psi}\psi}(p)
			=4m_\psi^2h^2k 
			\Biggl(& 
			(1-n_F(E_\psi-\mu))\frac{
			(2E_\psi-\I p_0)^2-6 E_\psi^2-k^2+2p_0^2+\vec{q}^{\,2}}
			{E_\psi^3(E_\psi^2-k^2-(E_\psi+\I p_0)^2+\vec{q}^{\,2})^2
			}\\
			&-n_F(E_\psi+\mu)\frac{
			(2E_\psi+\I p_0)^2-6 E_\psi^2-k^2+2p_0^2+\vec{q}^{\,2}}
			{E_\psi^3(E_\psi^2-k^2-(E_\psi-\I p_0)^2+\vec{q}^{\,2})^2
			}\\
			&-n_F(E_\psi-\mu)(1-n_F(E_\psi-\mu))\frac{
			1
			}{E_\psi^2(E_\psi^2-k^2-(E_\psi+\I p_0)^2+\vec{q}^{\,2})T
			}\\
			&-n_F(E_\psi+\mu)(1-n_F(E_\psi+\mu))\frac{
			1
			}{E_\psi^2(E_\psi^2-k^2-(E_\psi-\I p_0)^2+\vec{q}^{\,2})T
			}\\
			&+(1-n_F(\tilde{E}_\psi-\mu))\frac{
			2
			}{\tilde{E}_\psi(E_\psi^2-(\tilde{E}_\psi-\I p_0)^2)^2
			}
			-n_F(\tilde{E}_\psi+\mu)\frac{
			2
			}{\tilde{E}_\psi(E_\psi^2-(\tilde{E}_\psi+\I p_0)^2)^2
			}
			\Biggr)\\
			%%%%%%%%%%%%%%%%%%%%%%%%%%%%%%%%%%%%%%%%%%%%%%%%%%
		\end{split}
\end{equation}
\begin{equation}
\label{eq:J_pi_tilde}
		\begin{split}
			%%%%%%%%%%%%%%%%%%%%%%%%%%%%%%%%%%%%%%%%%%%%%%%%%%
			\tilde{J}^{(\pi)}_{k,\bar{\psi}\psi}(p) =-2h^2k\Biggl[&
			(1-n_F(E_\psi-\mu))\Biggl(\frac{
			2\I p_0 E_\psi^3
			(|\vec{q}|\cos\varphi-k)+4\I p_0 E_\psi k^2 (k-|\vec{q}|\cos\varphi)
			}{E_\psi^3k(E_\psi^2-k^2-(E_\psi+\I p_0)^2+\vec{q}^{\,2})^2
			}\\
			&\hspace{30mm}-\frac{
			k^2(p_0^2-k^2+\vec{q}^{\,2})(k-|\vec{q}|\cos\varphi)
			+E_\psi^2|\vec{q}|((k^2+p_0^2+\vec{q}^{\,2})\cos\varphi-2k|\vec{q}|)
			}{E_\psi^3k(E_\psi^2-k^2-(E_\psi+\I p_0)^2+\vec{q}^{\,2})^2
			}\Biggr)\\
			&+n_F(E_\psi+\mu)\Biggl(\frac{
			2\I p_0 E_\psi^3
			(|\vec{q}|\cos\varphi-k)+4\I p_0 E_\psi k^2 (k-|\vec{q}|\cos\varphi)
			}{E_\psi^3k(E_\psi^2-k^2-(E_\psi-\I p_0)^2+\vec{q}^{\,2})^2
			}\\
			&\hspace{25mm}+\frac{
			k^2(p_0^2-k^2+\vec{q}^{\,2})(k-|\vec{q}|\cos\varphi)
			+E_\psi^2|\vec{q}|((k^2+p_0^2+\vec{q}^{\,2})\cos\varphi-2k|\vec{q}|)
			}{E_\psi^3k(E_\psi^2-k^2-(E_\psi-\I p_0)^2+\vec{q}^{\,2})^2
			}\Biggr)\\
			&+n_F(E_\psi-\mu)(1-n_F(E_\psi-\mu))\frac{
			k(k-|\vec{q}|\cos\varphi)+\I p_0 E_\psi 
			}{E_\psi^2(E_\psi^2-k^2-(E_\psi+\I p_0)^2+\vec{q}^{\,2})
			T}\\
			&+n_F(E_\psi+\mu)(1-n_F(E_\psi+\mu))\frac{
			k(k-|\vec{q}|\cos\varphi)-\I p_0 E_\psi 
			}{E_\psi^2(E_\psi^2-k^2-(E_\psi-\I p_0)^2+\vec{q}^{\,2})
			T}\\
			&+(1-n_F(\tilde{E}_\psi-\mu))\frac{
			|\vec{q}|\cos\varphi
			(k^2-p_0^2+\vec{q}^{\,2}-2\I p_0\tilde{E}_\psi)
			-2k(\vec{q}^{\,2}-\I p_0\tilde{E}_\psi)
			}{k\tilde{E}_\psi(
			E_\psi^2-(\tilde{E}_\psi-\I p_0)^2)^2
			}\\
			&-n_F(\tilde{E}_\psi+\mu)\frac{
			|\vec{q}|\cos\varphi
			(k^2-p_0^2+\vec{q}^{\,2}+2\I p_0\tilde{E}_\psi)
			-2k(\vec{q}^{\,2}+\I p_0\tilde{E}_\psi)
			}{k\tilde{E}_\psi(
			E_\psi^2-(\tilde{E}_\psi+\I p_0)^2)^2
			}\Biggr]\\
			%%%%%%%%%%%%%%%%%%%%%%%%%%%%%%%%%%%%%%%%%%%%%%%%%%
		\end{split}
\end{equation}

\end{widetext}

\bibliography{qcd}

\end{document}